\documentclass[12pt,a4paper]{article}

\usepackage[a4paper,vmargin=2.5cm,hmargin=2cm]{geometry}
\usepackage{lmodern}
\usepackage[T1]{fontenc}
\usepackage{amsmath,amssymb}
\usepackage[bookmarks=true,hyperfigures=true]{hyperref}
\usepackage{graphicx} 
\usepackage[font=small,labelfont=bf,width=0.85\textwidth]{caption} 
\usepackage[nosort]{cite} 
\usepackage[bulletsep]{collref} 
\usepackage{fixmath} 
\usepackage{microtype}
\providecommand{\microtypesetup}[1]{}

\usepackage{fancybox}
\newcommand\eqbox[1]{\ovalbox{\ensuremath{\displaystyle\mathstrut#1}}}
\graphicspath{{./figures/}}

\makeatletter \let\@keywords\@empty \let\@subject\@empty
\providecommand{\keywords}[1]{\gdef\@keywords{#1}}
\providecommand{\subject}[1]{\gdef\@subject{#1}}
\def\thetitle{\@title}
\def\theauthor{\@author}
\def\thesubject{\@subject}
\def\thedate{\@date}
\def\thekeywords{\@keywords}
\makeatother
\AtBeginDocument{%
\hypersetup{pdftitle={\thetitle}}%
\hypersetup{pdfauthor={\theauthor}}%
\hypersetup{pdfsubject={\thesubject}}%
\hypersetup{pdfkeywords={\thekeywords}}%
}

\providecommand{\hypersetup}[1]{}
\providecommand{\texorpdfstring}[2]{#1}

\hypersetup{plainpages=false}
\hypersetup{pdfpagemode=UseNone}
\hypersetup{bookmarksnumbered=true}
\hypersetup{pdfstartview=FitH} 
\hypersetup{colorlinks=false}
\hypersetup{citebordercolor={.5 1 .5}}
\hypersetup{urlbordercolor={.5 1 1}}
\hypersetup{linkbordercolor={1 .7 .7}}

\numberwithin{equation}{section}

\let\oldbfseries=\bfseries
\let\oldmdseries=\mdseries
\let\oldnormalfont=\normalfont
\renewcommand{\bfseries}{\oldbfseries\boldmath}
\renewcommand{\mdseries}{\oldmdseries\unboldmath}
\renewcommand{\normalfont}{\oldnormalfont\unboldmath}

\newcommand{\sfrac}[2]{{\textstyle\frac{#1}{#2}}}
\newcommand{\half}{\sfrac{1}{2}}

\newcommand{\op}[1]{\mathcal{#1}}

\newcommand{\gen}[1]{\mathfrak{#1}}

\newcommand{\alg}[1]{\mathfrak{#1}}
\newcommand{\grp}[1]{\mathrm{#1}}
\newcommand{\superN}{\mathcal{N}}

\newcommand{\amp}{\mathcal{A}}
\newcommand{\grass}{\mathcal{G}}

\newcommand{\transpose}{^\mathsf{T}}

\newcommand{\cc}{z'}

\newcommand{\cA}{\ensuremath{\mathcal{A}}}
\newcommand{\cB}{\ensuremath{\mathcal{B}}}
\newcommand{\cC}{\ensuremath{\mathcal{C}}}
\newcommand{\cF}{\ensuremath{\mathcal{F}}}
\newcommand{\cN}{\ensuremath{\mathcal{N}}}
\newcommand{\cO}{\ensuremath{\mathcal{O}}}
\newcommand{\cU}{\ensuremath{\mathcal{U}}}
\newcommand{\cV}{\ensuremath{\mathcal{V}}}
\newcommand{\cW}{\ensuremath{\mathcal{W}}}
\newcommand{\cY}{\ensuremath{\mathcal{Y}}}
\newcommand{\cZ}{\ensuremath{\mathcal{Z}}}

\newcommand{\GL}{\grp{GL}}
\newcommand{\OG}{\grp{OG}}

\newcommand{\dd}{d}
\newcommand{\dL}{\dd^{2|3}\!\Lambda}
\newcommand{\eps}{\varepsilon}

\newcommand{\deltaP}{\ensuremath{\delta^3(P)}}
\newcommand{\deltaQ}{\ensuremath{\delta^6(Q)}}
\newcommand{\deltaPQ}{\ensuremath{\deltaP\,\deltaQ}}

\newcommand{\bfm}[1]{\boldsymbol{#1}}
\newcommand{\indup}[1]{_{\mathrm{#1}}}

\newcommand{\supup}[1]{^{\mathrm{#1}}}

\newcommand{\nn}{\nonumber}

\newcommand{\brk}[1]{(#1)}
\newcommand{\lrbrk}[1]{\left(#1\right)}
\newcommand{\bigbrk}[1]{\bigl(#1\bigr)}
\newcommand{\biggbrk}[1]{\biggl(#1\biggr)}
\newcommand{\Bigbrk}[1]{\Bigl(#1\Bigr)}
\newcommand{\sbrk}[1]{[#1]}
\newcommand{\lrsbrk}[1]{\left[#1\right]}
\newcommand{\bigsbrk}[1]{\bigl[#1\bigr]}

\newcommand{\Bigsbrk}[1]{\Bigl[#1\Bigr]}
\newcommand{\brc}[1]{\{#1\}}

\newcommand{\vev}[1]{\langle#1\rangle}

\newcommand{\comm}[2]{[#1,#2]}
\newcommand{\lrcomm}[2]{\left[#1,#2\right]}

\newcommand{\abs}[1]{|#1|}

\newcommand{\spaa}[1]{{\langle#1\rangle}}

\newcommand{\spbb}[1]{{[#1]}}

\makeatletter
\def\mr@ignsp#1 {\ifx\:#1\@empty\else #1\expandafter\mr@ignsp\fi}%
\newcommand{\multiref}[1]{\begingroup
\xdef\mr@no@sparg{\expandafter\mr@ignsp#1 \: }%
\def\mr@comma{}%
\@for\mr@refs:=\mr@no@sparg\do{\mr@comma\def\mr@comma{,}\ref{\mr@refs}}%
\endgroup}
\makeatother

\newcommand{\hypref}[2]{\ifx\href\asklfhas #2\else\href{#1}{#2}\fi}
\newcommand{\secref}[1]{Section~\multiref{#1}}
\newcommand{\appref}[1]{Appendix~\multiref{#1}}
\newcommand{\tabref}[1]{Table~\multiref{#1}}
\newcommand{\figref}[1]{Figure~\multiref{#1}}
\renewcommand{\eqref}[1]{(\multiref{#1})}

\makeatletter
\newlength{\apb@width}
\newcommand{\autoparbox}[2][c]{\settowidth{\apb@width}{#2}\parbox[#1]{\apb@width}{#2}}
\newcommand{\includegraphicsbox}[2][]{\autoparbox{\includegraphics[#1]{#2}}}
\makeatother

\usepackage{xcolor}

\title{Integrable Amplitude Deformations\texorpdfstring{\\}{ }for
\texorpdfstring{$\superN=4$}{N=4}
Super Yang--Mills and ABJM Theory}

\author{%
Till Bargheer,
Yu-tin Huang,
Florian Loebbert,
Masahito Yamazaki}

\begin{document}

\pdfbookmark[1]{Title Page}{title}

\thispagestyle{empty}

\begin{flushright}
\texttt{DESY 14-128}\\
\texttt{IPMU-14-0157}\\
\texttt{HU-EP-14/29}
\end{flushright}

\vfill

\begin{center}

\begingroup\Large\bfseries\thetitle\par\endgroup

\vfill

\begingroup\scshape
Till Bargheer$^{1,2}$,
Yu-tin Huang$^{1,3}$,\\
Florian Loebbert$^{1,4}$,
Masahito Yamazaki$^{1,5}$\par
\endgroup

\vfill

\begingroup\itshape
$^1$Institute for Advanced Study,
School of Natural Sciences,\\
1 Einstein Drive, Princeton, New Jersey 08540, USA

\bigskip

$^2$DESY Theory Group, DESY Hamburg,\\
Notkestra\ss e 85, D-22603 Hamburg, Germany

\bigskip

$^3$Department of Physics and Astronomy,
National Taiwan University,\\
Taipei 10617, Taiwan, ROC

\bigskip

$^4$Institut f\"ur Physik, Humboldt-Universit\"at zu Berlin, \\
Newtonstrasse 15, D-12489 Berlin, Germany

\bigskip

$^5$Kavli IPMU (WPI),
University of Tokyo,\\
Kashiwa, Chiba 277-8583, Japan
\par\endgroup

\bigskip

{\ttfamily
\href{mailto:bargheer@ias.edu}{bargheer@ias.edu},
\href{mailto:yutinyt@gmail.com}{yutinyt@gmail.com},\\
\href{mailto:loebbert@ias.edu}{loebbert@ias.edu},
\href{mailto:masahito.yamazaki@ipmu.jp}{masahito.yamazaki@ipmu.jp}}

\vfill

\textbf{Abstract}

\bigskip

\begin{minipage}{12cm}
We study Yangian-invariant
deformations of scattering amplitudes in
4d $\superN=4$ supersymmetric Yang--Mills theory and 3d
$\superN=6$ Aharony--Bergman--Jafferis--Maldacena (ABJM) theory.
In particular, we obtain the deformed Gra{\ss}mannian integral for 4d
$\superN=4$ super Yang--Mills theory, both in momentum and momentum-twistor space.
For 3d ABJM theory, we initiate the study of deformed scattering amplitudes.
We investigate general deformations of on-shell diagrams,
and find the deformed Gra{\ss}mannian integral for this
theory. We furthermore introduce the algebraic R-matrix
construction of deformed Yangian invariants for ABJM theory.
\end{minipage}

\end{center}

\vfill
\vfill

\setcounter{page}{0}


\newpage

\setcounter{tocdepth}{2}
\hrule height 0.75pt
\pdfbookmark[1]{\contentsname}{contents}
\microtypesetup{protrusion=false}
\tableofcontents
\microtypesetup{protrusion=true}
\vspace{0.8cm}
\hrule height 0.75pt
\vspace{1cm}

\section{Introduction}
\label{sec:intro}

Recently it has become clear that the physics of scattering amplitudes contains a plethora of
interesting mathematical structures and unexpected symmetries.%
\footnote{For example, see~\cite{Elvang:2013cua} for a recent review.}
The
prime examples for investigating these phenomena are $\superN=4$
super Yang--Mills and $\superN=6$ super Chern--Simons
(ABJM)~\cite{Aharony:2008ug, Hosomichi:2008jb} theory in four and three
dimensions, respectively. Both of these theories are believed to be equivalent
to an AdS/CFT-dual string theory, and both are believed to be completely
\emph{integrable} in the planar limit.
\medskip

In the context of scattering amplitudes, integrability is realized as
a Yangian symmetry acting on the external legs of the supersymmetric
amplitude~\cite{Drummond:2009fd}. Equivalently, the Yangian can be
formulated as the combination of superconformal and dual
superconformal symmetry, where---at least in four dimensions---the
latter arises from the duality between amplitudes and Wilson loops. The Yangian
symmetry is highly restrictive and, when combined with locality, completely
fixes the tree-level scattering matrix~\cite{Bargheer:2009qu,Korchemsky:2009hm}.
\medskip

Lately, it has been noticed that the tree-level S-matrix of
$\superN=4$ SYM theory can be identified with the maximally length-changing
contributions of the dilatation operator of the same theory~\cite{Zwiebel:2011bx}.
The simplest example of this map is the four-point
amplitude, which serves as an integral kernel for the celebrated one-loop
dilatation operator alias a super spin-chain Hamiltonian. Like the spin-chain
Hamiltonian, the four-point scattering amplitude can be obtained from an
R-matrix which depends on a spectral parameter $z$~\cite{Ferro:2012xw,Ferro:2013dga}.
Remarkably, this implies the existence of a deformation of the
tree-level four-point amplitude in the parameter~$z$. Importantly,
this deformation is still Yangian invariant. In fact, it is necessary
to consider a corresponding deformation of the Yangian generators
known as the evaluation representation. This representation is more
general than the previously considered representation and thus allows
for more general invariant functions of the kinematical scattering
data. Interestingly, if the evaluation parameters are all real, the
deformed Yangian generators preserve the positive Gra{\ss}mannian.
\medskip

It turns out that such deformed invariants also exist for higher
multiplicities. The most natural approaches to construct these invariants are
closely related to the on-shell methods of Arkani-Hamed
\emph{et al.}\ (see e.g.~\cite{ArkaniHamed:2012nw}).
In four dimensions, in particular a diagrammatic
approach has been studied~\cite{Ferro:2012xw,Ferro:2013dga,Beisert:2014qba}, as
well as an R-matrix construction of Yangian invariants, similar in spirit to
the algebraic Bethe
ansatz~\cite{Chicherin:2013ora,Frassek:2013xza,Kanning:2014maa,Broedel:2014pia,Broedel:2014hca}.
\medskip

As these invariants are functions of the external data on which the S-matrix is
defined, a natural question is how they are related to the scattering
amplitudes. Preliminary attempts to relate these invariants to the
Britto--Cachazo--Feng--Witten (BCFW)~\cite{Britto:2004ap, Britto:2005fq} building blocks of scattering
amplitudes via a uniform set of deformation parameters, however, appear to
break down when going beyond six points and the
maximally-helicity-violating (MHV) level~\cite{Beisert:2014qba}.
More precisely, attempts to simultaneously deform all contributing
BCFW terms in a consistent fashion have failed so far.
\medskip

To circumvent the above difficulties, instead of deforming the individual BCFW
contributions, one might consider embedding the BCFW terms into a parent
integral with some unspecified contour, and which depends on the deformation
parameters. As one turns off the deformation, we are allowed to choose the
contour such that the integral reduces to the individual BCFW terms. In this
way, as the deformation parameters are introduced at the level of the parent
integral, there is \emph{a priori} no inconsistency. Luckily such an integral already
exists in the form of the Gra{\ss}mannian integral~\cite{ArkaniHamed:2009dn},
and the first task is to introduce deformations such that Yangian invariance is
preserved.
\medskip

In the present paper, we study deformed scattering amplitudes in
four-dimensional \mbox{$\superN=4$} super Yang--Mills theory, and in
three-dimensional $\superN=6$ super Chern--Simons theory (ABJM).
The purpose of this work is the following:

\begin{itemize}
\item[$\bullet$ 4d:]
We present a deformed and Yangian-invariant Gra{\ss}mannian integral and
discuss its r\^ole for further investigations of amplitude deformations.
We review and summarize the recent progress on deformed scattering amplitudes
in $\superN=4$ SYM theory in a compact form and highlight the connections among
different approaches.

\item[$\bullet$ 3d:]
In four dimensions, the deformation parameters are to some extent
associated with central charges or deformed helicities of the external
particles. The $\alg{osp}(6|4)$ algebra of ABJM theory does not contain a
central charge and the considered three-dimensional particles do not carry
helicity degrees of freedom. Thus it is interesting to ask whether or not
integrable deformations for ABJM theory exist as well.
Indeed, we find that the four-point amplitude allows for a one-parameter
deformation which is invariant under the evaluation representation of the
Yangian algebra $Y[\alg{osp}(6|4)]$. This deformed four-point vertex furnishes
the building block for invariants with higher multiplicities, which we
construct along the lines of the on-shell diagram methods
of~\cite{ArkaniHamed:2012nw, Huang:2013owa}. We then propose a deformation of
the orthogonal Gra{\ss}mannian integral introduced in~\cite{Lee:2010du} and
show that it is consistent with the previous investigations. Finally, we also
introduce an algebraic R-matrix construction of deformed Yangian invariants for
the three-dimensional theory.
\end{itemize}

This paper is organized as follows.
In \secref{sec:intdefs4d} we review the construction of deformed
Yangian invariants in $\superN=4$ SYM theory: In particular, we
present the deformed Gra{\ss}mannian integral in
\secref{sec:defgrass}, whose Yangian invariance follows from the
on-shell diagram formalism discussed in \secref{sec:defon4d} or the
direct proof in \appref{app:4dgrassinv}. We also obtain the deformed
momentum-twistor version of the Gra{\ss}mannian.
The study of deformed scattering amplitudes in three-dimensional ABJM
theory is initiated in \secref{sec:ABJM}: We demonstrate
Yangian invariance of the deformed four-point amplitude in
\secref{sec:4ptabjm}. We then propose a deformed orthogonal
Gra{\ss}mannian integral and show its Yangian symmetry in
\secref{sec:defgrass3d}. We explain how to build deformed
Yangian-invariant on-shell diagrams in \secref{sec:glue}, and introduce an
algebraic R-matrix construction of these invariants in \secref{sec:RmatABJM}.
Finally, we comment on differences and similarities between the four- and
three-dimensional case and point out interesting directions for the future in
\secref{sec:discussion}.

\section{Integrable Deformations in \texorpdfstring{$\superN=4$}{N=4} SYM Theory}
\label{sec:intdefs4d}

Recently, it has been found that on-shell diagrams allow for
interesting deformations that maintain the complete Yangian
invariance~\cite{Ferro:2012xw,Ferro:2013dga,Beisert:2014qba}. Here, we
will briefly summarize these ideas in preparation for the ABJM case
below, and comment on a number of features of the deformations.

The
Yangian level-one generators generically take the form%
\footnote{More about Yangian symmetry in the present context can be found
in~\cite{Drummond:2009fd,Bargheer:2010hn,Beisert:2010jq}. For general
introductions, see~\cite{Bernard:1993ya,MacKay:2004tc}.}
\begin{equation}
\gen{\widehat J}^a
=
f^a{}_{bc}
\sum_{\substack{i,j=1\\i<j}}^n
\gen{J}^b_i\,
\gen{J}^c_j
+
\sum_{i=1}^n
u_i\,\gen{J}^a_i\,,
\label{eq:lev1_J}
\end{equation}
where $\gen{J}_i^a$ are the local level-zero generators
acting on the leg $i$, and $f^a{}_{bc}$ are the structure constants of the
level-zero algebra. The evaluation parameters $u_i$ are set to zero in
the undeformed case. For the superconformal symmetry algebra
$\alg{psu}(2,2|4)$ in twistor variables,%
\footnote{For a definition of the twistor variables
$\cZ_i$, see~\cite{Witten:2003nn}. In $(2,2)$ signature,
$\cZ_i^{\cA}=(\tilde\mu_i^{\dot\alpha},\tilde\lambda_i^{\dot\alpha}|\eta_i^A)$, where
$\tilde\mu_i$ is the Fourier transform of the momentum spinor
$\lambda_i$. Here,
$p_i^\mu=\sigma^\mu_{\alpha\dot\alpha}\lambda_i^\alpha\tilde\lambda_i^{\dot\alpha}$,
and the anticommuting spinor $\eta_i^A$, with $A=1,\dots,4$, parametrizes the
$\superN=4$ on-shell superfield~\cite{Nair:1988bq}.}
the level-zero and level-one generators
take the form~\cite{Drummond:2009fd}
\begin{align}
\gen{J}^{\cA}{}_{\cB}
&=
\sum_{i=1}^n
\gen{J}_i^{\cA}{}_{\cB}
\,,
\qquad
\gen{J}_i^{\cA}{}_{\cB}
=
\cZ_i^{\cA}\frac{\partial}{\partial\cZ_i^{\cB}}
-\text{(trace)}\,,
\label{eq:lev0}
\\
\gen{\widehat J}^\cA{}_\cB
&=\sum_{i<j}(-1)^\cC\lrsbrk{\gen{J}_i^\cA{}_\cC\gen{J}_j^\cC{}_\cB-(i\leftrightarrow j)}
+\sum_i u_i
\gen{J}_i^\cA{}_\cB
\,,
\label{eq:lev1_Z}
\end{align}
where $\cA$, $\cB$ are fundamental $\alg{su}(2,2|4)$
indices.
The central charge operators read
\begin{equation}
\gen{C}_i
=
-\cZ_i^{\cC}\frac{\partial}{\partial\cZ_i^\cC}\,.
\label{eq:centop}
\end{equation}

\subsection{Deformed On-Shell Diagrams}
\label{sec:defon4d}

Every on-shell diagram is either a BCFW term of a tree amplitude or
loop integrand, or a leading singularity of a loop amplitude~\cite{ArkaniHamed:2012nw}.
In the following, we review the known integrable deformations of
general on-shell diagrams.

\paragraph{Three-Vertices.}

The basic building blocks for the deformed on-shell diagrams are the
three-point vertices
\begin{equation}
\widehat\amp_3^{\circ}
=\!\!\!\includegraphicsbox[scale=1]{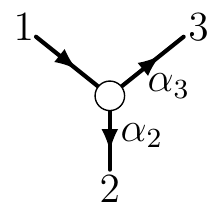}\!\!
=\int
\frac{\dd\alpha_2}{\alpha_2^{1+a_2}}
\frac{\dd\alpha_3}{\alpha_3^{1+a_3}}\,
\delta^{4|4}(C_{\circ}\cdot\cZ)\,,
\quad
\widehat\amp_3^{\bullet}
=\!\!\!\includegraphicsbox[scale=1]{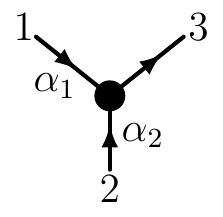}\!\!
=\int
\frac{\dd\alpha_1}{\alpha_1^{1+a_1}}
\frac{\dd\alpha_2}{\alpha_2^{1+a_2}}\,
\delta^{8|8}(C_{\bullet}\cdot\cZ)\,,
\label{eq:3vert}
\end{equation}
where
\begin{equation}
C_{\circ}=
\begin{pmatrix}
1 & \alpha_2 & \alpha_3
\end{pmatrix}\,,
\qquad
C_{\bullet}=
\begin{pmatrix}
1 & 0 & \alpha_1\\
0 & 1 & \alpha_2
\end{pmatrix}\,,
\end{equation}
and $\cZ_i^A$ are twistor variables that parametrize the
external states.
Unlike the undeformed vertices (with $a_i=0$), these
vertices have non-vanishing eigenvalues $c_i$ under the
action of the ``local'' central charges \eqref{eq:centop},
where
\begin{align}
\widehat\amp_3^{\circ}:
&&
c_1&=a_2+a_3\equiv a_1\,,&
c_2&=-a_2\,,&
c_3&=-a_3\,,
\nn\\
\widehat\amp_3^{\bullet}:
&&
c_1&=a_1\,,&
c_2&=a_2\,,&
c_3&=-a_1-a_2\equiv-a_3\,.
\label{eq:cofa}
\end{align}
They are invariant under the Yangian with evaluation parameters
$u_i$, where~\cite{Beisert:2014qba}%
\footnote{An overall shift of all evaluation parameters $u_i$ is
generated by the level-zero symmetry and is thus trivial, hence all
relevant parameters are captured by the differences of consecutive
evaluation parameters.}
\begin{align}
\widehat\amp_3^{\circ}:
&&
c_1&=u_3-u_2\,,&
c_2&=u_1-u_3\,,&
c_3&=u_2-u_1\,,
\nn\\
\widehat\amp_3^{\bullet}:
&&
c_1&=u_2-u_3\,,&
c_2&=u_3-u_1\,,&
c_3&=u_1-u_2\,.
\label{eq:3ptinvcond}
\end{align}
Converting the twistors back to spinor-helicity variables, the
three-vertices evaluate to the deformed amplitudes
\begin{equation}
\widehat\amp_3^{\circ}
=
\frac{\delta^4(P)\,\delta^4(\tilde Q)}{\spbb{12}^{1+a_3}\spbb{23}^{1-a_2-a_3}\spbb{31}^{1+a_2}}\,,
\qquad
\widehat\amp_3^{\bullet}
=
\frac{\delta^4(P)\,\delta^8(Q)}{\spaa{12}^{1-a_1-a_2}\spaa{23}^{1+a_1}\spaa{31}^{1+a_2}}\,.
\end{equation}
Here,
$\spbb{ij}\equiv
\varepsilon_{\dot\alpha\dot\beta}
\tilde{\lambda}_i^{\dot{\alpha}}
\tilde{\lambda}_{j}^{\dot{\beta}}$,
$\spaa{ij}\equiv
\varepsilon_{\alpha\beta}\lambda_i^\alpha\lambda_j^{\beta}$,
$P\equiv\sum_{i=1}^n\lambda_i\tilde{\lambda}_i$,
$Q\equiv\sum_{i=1}^n\lambda_i\eta_i$,
and
$\tilde{Q}\equiv(\spbb{12}\eta_3+\spbb{23}\eta_1+\spbb{31}\eta_2)$.

\paragraph{Gluing.}

All bigger on-shell diagrams $\cY$ can be built by iterating
two gluing operations: Taking products,
\begin{equation}
(\cY_1,\cY_2)\mapsto\cY_1\cY_2\,,
\end{equation}
and fusing lines,
\begin{equation}
\cY(\cZ_1,\dots,\cZ_n,\cZ_I,\cZ_J)
\mapsto
\int\dd^{3|4}\cZ_I\,
\cY(\cZ_1,\dots,\cZ_n,\cZ_I,\cZ_J)\big|_{\cZ_J=\cZ_I^-}\,,
\label{eq:fusion}
\end{equation}
where $\cZ_I^-$ is the twistor of line $I$ with inverse
momentum. Yangian invariance is preserved under both of these
operations. In addition to the twistor, each external line of a deformed diagram
carries two labels: A central charge $c_i$ and an evaluation parameter
$u_i$. When fusing lines of deformed diagrams, Yangian invariance
requires that~\cite{Beisert:2014qba}
\begin{equation}
c_I=-c_J\,,
\qquad
u_I=u_J\,.
\label{eq:4dglucon}
\end{equation}
Successively applying these two operations generates all
Yangian-invariant deformed on-shell diagrams. Combining \eqref{eq:4dglucon}
with the invariance conditions \eqref{eq:3ptinvcond} for the
three-vertices, one finds that the external central charges $c_i$ and
the evaluation parameters $u_i$ of any invariant diagram must obey~\cite{Beisert:2014qba}
\begin{equation}
\eqbox{u^+_i=u^-_{\sigma(i)}\,,}
\label{eq:invcond}
\end{equation}
where
\begin{align}
u_i^{\pm}\equiv u_i\pm c_i\,,
\end{align}
and $\sigma$ is the permutation
associated to the diagram. The permutation is obtained as
follows: Starting at the external line $i$, follow a path through the
diagram, turning left/right at each white/black vertex. The external
line at the end of the path will be $\sigma(i)$.%
\footnote{While every on-shell diagram has a unique permutation
associated to it, the converse is only true for ``reduced'' diagrams.
See~\cite{ArkaniHamed:2012nw} for more details.}
%

\paragraph{General Deformed Diagrams.}

Every deformed on-shell diagram can be written as
\begin{equation}
\widehat\cY(1,\dots,n)
=
\int\prod_{j=1}^{n\indup{F}-1}\frac{\dd\alpha_j}{\alpha_j^{1+a_j}}\,
\delta^{4k|4k}(C\cdot\cZ)\,,
\label{eq:edgedef}
\end{equation}
where $n\indup{F}$ is the number of faces of the diagram, $\alpha_j$
are a minimal number of edge variables,%
\footnote{The edge variables are essentially BCFW shift parameters.
Fixing a $\grp{GL(1)}$-gauge redundancy at each vertex, their number
can always be reduced to $n\indup{F}-1$.}
and $C$ is the matrix constructed from
the edge variables by ``boundary measurement'' as explained in~\cite{ArkaniHamed:2012nw}.
The gluing conditions \eqref{eq:4dglucon}
together with the identifications \eqref{eq:cofa} for the
three-vertices imply that the deformation parameter $a_j$ in the
exponent of an edge variable $\alpha_j$ equals the central charge
on the respective line up to a sign:
\begin{equation}
c_i=\pm a_i\,.
\end{equation}
Here the sign is plus/minus if the arrow on the line points in
the same/opposite direction as the permutation path.

The constraints \eqref{eq:invcond} impose $n$ conditions on the $2n$
central charges and evaluation parameters; hence, it is clear that
\emph{every} diagram admits $n-1$ independent non-trivial deformation
parameters (an additional parameter is the trivial uniform
shift of all $u_i$).

\paragraph{R-matrix Construction.}

Alternatively, (deformed) on-shell amplitudes can be constructed by
acting with a chain of R-matrices on a suitable ``vacuum state,''
where the action of the R-matrices exactly corresponds to the
insertion of a (deformed) BCFW
bridge~\cite{Chicherin:2013ora,Frassek:2013xza,Kanning:2014maa,Broedel:2014pia}.
Each R-matrix\,/\,BCFW
bridge contributes an adjacent transposition $\sigma_i$ to the (decorated) permutation $\sigma$
associated with the final diagram.
Following the procedure of~\cite{ArkaniHamed:2012nw},
we can associate a canonical decomposition into transpositions
$\sigma_i$,
\begin{align}
\sigma=\sigma_\ell\dots\sigma_1\,.
\label{sigmadecomp}
\end{align}
The deformed amplitude \eqref{eq:edgedef}
can then be written as
\begin{align}
\widehat{\cY}(\vec{u})=R_{\sigma_\ell}(a_\ell)\dots R_{\sigma_1}(a_1)\,
\delta^{4k|4k}(C_{\rm vac}\cdot\cZ)\,,
\label{Arecursive}
\end{align}
Here $C_{\rm vac}$ is a suitable ``vacuum matrix,'' which is a $k\times n$
matrix with $k$
unit columns and $(n-k)$ zero columns.
For example, for $n=6$, $k=3$, a valid choice is
\begin{align}
C_{\rm vac}=
\begin{pmatrix}
1 & 0 & 0 & 0 & 0 & 0 \\
0 & 0 & 1 & 0 & 0 & 0 \\
0 & 0 & 0 & 0 & 0 & 1
\end{pmatrix}
\,.
\label{CvacEg}
\end{align}
The operator $R_{\sigma=(ij)}$ is defined by~\cite{Chicherin:2013ora}
\begin{align}
\label{Rdef}
R_{\sigma=(ij)}(a) f(\cZ)
=\int\frac{d\alpha}{\alpha^{1+a}}
\,f(\cZ)\Big|_{\cZ_i\to\cZ_i+\alpha\cZ_j}
\,.
\end{align}
This R-operator can be identified with (an integral kernel for) an
R-matrix; the shift of $\cZ$ in the definition is nothing but the
BCFW shift. We can prove \eqref{Arecursive} by induction with respect to the
number of BCFW bridges.

This algebraic formulation is another way of demonstrating the Yangian invariance of the amplitude
(at the level of on-shell diagrams), and the integrable structures behind
it. In \secref{sec:RmatABJM} we provide a similar discussion for the ABJM theory.

\subsection{Deformed Gra{\ss}mannian Integral}
\label{sec:defgrass}

In this section we present the deformed Gra{\ss}mannian integral%
\footnote{Note that the deformed Gra{\ss}mannian formula as well as
its momentum-twistor version discussed below, were independently
obtained in~\cite{Ferro:2014gca}.}
for $\superN=4$ SYM theory as a special case of the above on-shell diagrams.
Being embedded into the on-shell-diagram formalism already implies the
Yangian symmetry of the deformed integral. We additionally demonstrate
the Yangian invariance of the Gra{\ss}mannian integral explicitly in \appref{app:4dgrassinv}.

\paragraph{The Gra{\ss}mannian Integral.}

A special class of diagrams are the ``top cells''~\cite{ArkaniHamed:2009dn}. These are diagrams
of maximal dimension (maximal number of integration variables). Their
name stems from the fact that all lower-dimensional on-shell diagrams
are realized as (iterated) boundaries of top-cell diagrams. They can
be classified by the number $n$ of external lines and the helicity
\begin{equation}
k=2n\indup{b}+n\indup{w}-n\indup{i}\,,
\end{equation}
where $n\indup{b/w}$ is the number of black/white vertices, and
$n\indup{i}$ is the number of internal lines. For each $n$ and $k$,
there is a unique top-cell diagram. It is the reduced diagram with the
maximal number of faces, $n\indup{F}=k(n-k)+1$. Every boundary
measurement on the top-cell diagram equals a gauge-fixed version of the
Gra{\ss}mannian integral of~\cite{ArkaniHamed:2009dn},
\begin{equation}
\grass_{n,k}(\cZ_1,\dots,\cZ_n)
=
\int\frac{\dd^{k\cdot n}C}{\abs{\grp{GL}(k)}}
\frac{1}{M_1^{1+b_1}\dots M_n^{1+b_n}}
\delta^{4k|4k}(C\cdot\cZ)\,,
\label{eq:topcell}
\end{equation}
where $M_i=\abs{i,\dots,i+k-1}$ is the $i$'th minor of $C$. The
integrand is invariant under $C\mapsto\grp{GL}(k)\cdot C$, and
$\abs{\grp{GL}(k)}$ is the volume of the gauge group. The permutation
associated to the top cell simply is a $k$-fold cyclic shift,
$\sigma:\brc{1,\dots,n}\mapsto\brc{n-k+1,\dots,n,1,\dots,n-k}$.
Noting that $c_i=-(b_{i-k+1}+\dots+b_i)$, the invariance conditions
\eqref{eq:invcond} imply
\begin{equation}
\eqbox{b_i=\half(u_i^--u_{i-1}^-)=\half(u_{i-k}^+-u_{i-k-1}^+)\,,}
\label{eq:topcellexp}
\end{equation}
and hence $\sum_i b_i=0$, which ensures $\grp{GL}(k)$ invariance.
The Yangian invariance of \eqref{eq:topcell}, with the parameters set
to \eqref{eq:topcellexp}, can also be
shown by directly acting with the Yangian generators,
see~\appref{app:4dgrassinv}.

\paragraph{Singularities and Residues.}

In the undeformed case, lower-dimensional on-shell diagrams are
obtained from the top cell \eqref{eq:topcell} by localizing some of
the integrations on residues. The $n$-point, helicity $k$ top-cell
diagram is defined in terms of $k(n-k)$ integrations, of which $2n-4$
can be performed trivially, using the bosonic delta functions.
Iteratively localizing all
remaining integrations on a suitable combination of residues gives the
tree-level amplitude $\amp_{n,k}$.
In terms of edge variables, taking a residue amounts to setting one
edge variable to zero, and the residue is given by the on-shell
diagram with the corresponding edge removed. Hence tree-level
amplitudes are given by summing a certain set of on-shell diagrams.

In the presence of generic deformations, the integrations can no
longer be performed on residues. From the perspective of the
gauge-fixed integral \eqref{eq:edgedef}, one possibility to proceed is to
just evaluate the integral on the same contour as in the undeformed
case. This requires us to set the deformation parameters $a_j$ to zero on the
respective edge variables, as otherwise the contour would not be
closed, due to branch cuts. In this case, the result of the integration is a sum
over the same set of on-shell diagrams as in the undeformed case,
where now each diagram is deformed. However, setting some of the
parameters $a_j$ to zero reduces the space of deformation moduli.
In fact, it was noted in~\cite{Beisert:2014qba} that for generic
tree amplitudes, the constraints imposed by this procedure rule out
all deformations. In other words, the integral can only be localized
by residues on a standard tree contour when \emph{all} moduli
$a_j$ are set to zero.
Setting the exponent of an edge variable $\alpha$ to unity and
localizing the integration by residue on $\alpha=0$ amounts to
``undoing'' a deformed BCFW bridge (R-matrix insertion). The
deformation parameters of the diagram
obtained in this way from the top cell will by construction satisfy
\eqref{eq:invcond} for $\sigma$ being the $k$-fold cyclic shift of the
top cell; but they will also satisfy \eqref{eq:invcond} for all
intermediate permutations $\sigma'$ that lead from $\sigma$ to the permutation of the
final diagram. Therefore, localizing the Gra{\ss}mannian integral
generates only a subspace of all admitted deformations for all
lower-dimensional diagrams.

Another perspective on the incompatibility of BCFW and deformations is
provided by the Gra{\ss}mannian integral in its original,
un-gauge-fixed form \eqref{eq:topcell}. In the undeformed case, the tree-level amplitudes
are given by the residues as the integral localizes on the zeros of
the minors. Thus an amplitude is identified with the locus of
zeros for a collection of minors. To further admit this localization, the exponents
of these minors should be undeformed. As the number of BCFW terms
increases, eventually this collection of minors covers the whole set,
and thus no deformation is allowed. Indeed
from~\cite{ArkaniHamed:2009dn}, we see that in the seven-point NMHV
case, the
collection of minors involved in the localization covers six of them,
and since the sum of $b_i$'s must vanish, there are no admissible
deformations left.

An alternative and perhaps more promising treatment for the deformed Gra{\ss}mannian integral
\eqref{eq:topcell} would be to leave the deformation parameters
generic and to evaluate the integral by other means on
an appropriate contour. We will comment on this idea in
\secref{sec:discussion} below.

\paragraph{Note on Positivity.}

There exists a remarkable relation between on-shell diagrams and
the positroid stratification of the
Gra{\ss}mannian~\cite{ArkaniHamed:2012nw}. The positroid stratification is
the classification of all distinct linear dependencies of consecutive
columns in the $C$-matrix, and it turns out that there is a one-to-one
correspondence between inequivalent on-shell diagrams and
inequivalent cells in the stratification. An interesting
property that can be associated with these cells is that there exist
parametrizations such that all non-vanishing minors are positive.
It has been noted that the Yangian generators generate diffeomorphisms
that act on the Gra{\ss}mannian in such a way that positivity is
preserved~\cite{ArkaniHamed:2012nw}.
One may ask if the deformed Yangian generators can still be
understood as positivity-preserving diffeomorphism. This is indeed the
case: The level-one generators are deformed by terms
$u_i\,\cZ_i\,\partial/\partial\cZ_i$, which, when acting on
the delta functions $\delta^{4k|4k}(C\cdot\cZ)$, translate
into $u_i\,\sum_a C_{a i}\,\partial/\partial C_{a i}$, which is
nothing but a little group scaling that simply rescales the $i$'th
column of the matrix $C$. Thus, as long as all evaluation parameters
$u_i$ are real, positivity of the cell is preserved.

\subsection{Relation to Deformed Momentum-Twistor Invariants}
\label{sec:defmomtwi}

Everything that has been stated above for on-shell diagrams in twistor
variables $\cZ$ is equally true for on-shell diagrams in
momentum-twistor variables~$\cW$: Replacing
$\cZ_i$ with $\cW_i$ in the expressions
\eqref{eq:edgedef,eq:3vert,eq:fusion,eq:topcell}, the resulting
momentum-twistor
diagrams are invariant under the momentum-twistor Yangian%
\footnote{Here, $\cW_i\equiv (\lambda^\alpha_i,
\mu_i^{\dot{\alpha}}|\chi_i^I)$, with
\begin{equation}
\mu_i^{\dot{\alpha}}\equiv\varepsilon_{\alpha\beta}\,y_{i}^{\dot{\alpha}\alpha}\lambda_i^{\beta},\quad
\chi_i\equiv\varepsilon_{\alpha\beta}\,\theta_{i}^{\alpha}\lambda_i^{\beta}\,,
\end{equation}
where the dual coordinates ($y_i,\theta_i$) are defined through
$y_i-y_{i+1}=p_i$ and $\theta_{i}-\theta_{i+1}=\lambda_i\eta_i$.}
with
generators~\cite{Drummond:2010qh}
\begin{gather}
\gen{\widehat I}^a
=
f^a{}_{bc}
\sum_{\substack{i,j=1\\i<j}}^n
\gen{I}^b_i\,
\gen{I}^c_j
+
\sum_{i=1}^n
v_i\,\gen{I}^a_i\,,
\label{eq:Wlev1}
\\
\gen{I}^{\cA}{}_{\cB}
=
\sum_{i=1}^n
\gen{I}_i^{\cA}{}_{\cB}
\,,
\qquad
\gen{I}_i^{\cA}{}_{\cB}
=
\cW_i^{\cA}\frac{\partial}{\partial\cW_i^{\cB}}
-\text{(trace)}\,,
\qquad
\gen{C}_i
=
\cW_i^{\cC}\frac{\partial}{\partial\cW_i^C}\,.
\label{eq:Wlev0}
\end{gather}
On the other hand, in the undeformed case it is known that for any
invariant $\cY(\cW)$ of the momentum-twistor Yangian, the
expression
\begin{equation}
\frac{\delta^4(P)\delta^8(Q)}{\spaa{12}\spaa{23}\dots\spaa{n1}}\,\cY(\cW)\,,
\label{eq:mhvtreexYW}
\end{equation}
when transformed to twistor variables $\cZ$, is an invariant
of the twistor-variable Yangian, and vice versa~\cite{Mason:2009qx,Drummond:2010qh}.%
\footnote{The underlying reason is that the Yangian of the ordinary
superconformal algebra with generators \eqref{eq:lev0,eq:lev1_Z} and the Yangian
of the dual superconformal algebra with generators
\eqref{eq:Wlev1,eq:Wlev0} in fact can be
mapped to each other~\cite{Drummond:2009fd,Drummond:2010qh}.}
It turns out that a
similar statement holds in the deformed case. This can be seen as
follows: An explicit procedure for reducing any on-shell twistor
diagram to a corresponding momentum-twistor diagram is given in
Section~8.3 of~\cite{ArkaniHamed:2012nw}. The deformation only
affects the integration measure, and hence the reduction procedure
applies in exactly the same way to the
deformed diagrams. Under the reduction, the minors of the matrix $C$
transform to minors of the reduced matrix $\tilde C$ as
\begin{equation}
\det C\big|_{\brk{i,\dots,i+k-1}}
=
\spaa{i,i+1}\spaa{i+1,i+2}\dots\spaa{i+k-2,i+k-1}\det{\tilde C}\big|_{\brk{i+1,\dots,i+k-2}}\,.
\end{equation}
Hence the only modification in the deformed case is a
deformation of the MHV tree prefactor in \eqref{eq:mhvtreexYW}.
In particular, for the deformed top cell \eqref{eq:topcell}, one finds
\begin{gather}
\widehat{\grass}_{n,k}(\cZ)
\xrightarrow{\cZ\to\cW}
\frac{\delta^4(P)\delta^8(Q)}{\spaa{12}^{1+(u_1^--u_{n-k+2}^-)/2}\dots\spaa{n1}^{1+(u_n^--u_{n-k+1}^-)/2}}
\,\widehat\grass_{n,k-2}(\cW)\,,
\label{eq:ZWreduct}
\\
\widehat\grass_{n,k}(\cW)
\equiv
\int\frac{\dd^{k\cdot n}\tilde C}{\abs{\grp{GL}(k)}}
\frac{1}{\tilde M_1^{1+b_n}\tilde M_2^{1+b_1}\dots\tilde M_n^{1+b_{n-1}}}
\delta^{4k|4k}(\tilde C\cdot\cW)\,,
\label{eq:dtopcell}
\end{gather}
where the $\tilde M_i$ are the minors of the reduced matrix
$\tilde C$.
At the same time, we know that the momentum-twistor top cell
$\widehat\grass_{n,k-2}(\cW)$ by itself is invariant under the
deformed momentum-twistor Yangian \eqref{eq:Wlev0,eq:Wlev1} once we
identify
\begin{equation}
\hat b_i\equiv b_{i-1}=\half(v_i^--v_{i-1}^-)\,,
\qquad
v_i^-=v_i-c\supup{dual}_i\,,
\end{equation}
according to \eqref{eq:topcellexp}.
Here, $c\supup{dual}_i=c_{i-1}$ are the eigenvalues of $\widehat\grass_{n,k-2}(\cW)$
under the local central charges
$\gen{C}_i$ in \eqref{eq:Wlev0}.
It follows that the deformed Gra{\ss}mannian integral is invariant,
both under the deformed original-twistor Yangian
\eqref{eq:lev1_Z,eq:lev0} and the deformed momentum-twistor Yangian
\eqref{eq:Wlev1,eq:Wlev0}, once one identifies
\begin{equation}
\eqbox{u_i^--u_{i-1}^-=v_{i+1}^--v_i^-\,,}
\end{equation}
where the prefactor in \eqref{eq:ZWreduct} has to be taken into
account in the invariance statement.
The relation \eqref{eq:ZWreduct} between
invariants of the twistor Yangian and the momentum-twistor Yangian
generalizes to all deformed on-shell diagrams that can be obtained
from the deformed Gra{\ss}mannian formula by localization on residues
as explained above:
\begin{equation}
\widehat{\cY}_{n,k}(\cZ)
\xrightarrow{\cZ\to\cW}
\frac{\delta^4(P)\delta^8(Q)}{\spaa{12}^{1+(u_1^--u_{n-k+2}^-)/2}\dots\spaa{n1}^{1+(u_n^--u_{n-k+1}^-)/2}}
\,\widehat\cY_{n,k-2}(\cW)\,.
\label{eq:eqiv}
\end{equation}
Here the parameters $u_i^-$ now satisfy additional constraints
imposed by setting the appropriate moduli $b_i$ to zero for the
purpose of localizing the integrations. It would be interesting to
understand whether the equivalence \eqref{eq:eqiv} extends also to
deformed diagrams that \emph{cannot} be obtained by localizing the
top cell, i.e.\ whose deformation parameters $u_i^-$ are
unconstrained. Also, it would be interesting to check whether the
deformed Yangians \eqref{eq:lev1_Z,eq:lev0} and
\eqref{eq:Wlev1,eq:Wlev0} still map to each other as in the undeformed
case.

\subsection{Examples}

Let us summarize the construction of invariant deformed diagrams, and
close the discussion of $\superN=4$ SYM deformations by
commenting on some interesting examples, including curious
deformations for MHV amplitudes as well as an explicit exposition of
the deformed six-point NMHV Gra{\ss}mannian integral.

\paragraph{Summary of Construction.}

Working out the admissible deformations for any given single diagram
works as follows~\cite{Beisert:2014qba}. First pick a perfect orientation and a set of
$(n\indup{F}-1)$ edge variables. The candidate invariant then is
\eqref{eq:edgedef}, and the invariance constraints on the external central charges
$c_{1,\dots,n}$, evaluation parameters $u_{1,\dots,n}$, and edge
variable parameters $a_{1,\dots,n\indup{F}-1}$ are the following: For each left-right
path%
\footnote{Turn left at each white vertex ($\overline{\text{MHV}}$),
turn right at each
black vertex (MHV).}
from site $i$ to site $j$, set $u^+_i=u^-_j$, which also must equal
the \emph{internal} parameter $u\indup{int}\pm c\indup{int}$ on each
labeled edge along the path. Here the sign depends on the direction
of the edge, and the internal central charge $c\indup{int}$ equals the
edge parameter $a_\ell$ on that edge, as follows from
\eqref{eq:cofa}.

For diagrams that appear in BCFW decompositions of tree-level
amplitudes, the number of integrations
$(n\indup{F}-1)$ equals the number $(2n-4)$ of bosonic delta
functions. In this case, the
expression for the diagram in terms of spinor-helicity
variables can be worked out by solving the delta-function constraints
for the edge variables, taking into account the resulting Jacobi factor.
In the case of MHV tree amplitudes (which are single top-cell diagrams),
the minors $(ij)$ of the matrix $C$ equal the spinor brackets
$\spaa{ij}$, and the edge variables are given in terms of these minors
by the function \texttt{bridgeToMinors} of the \texttt{Mathematica} package
given in~\cite{Bourjaily:2012gy}. Comparing the measure
$\brk{\alpha_1\dots\alpha_{2n-4}}^{-1}$ to the Parke--Taylor MHV
denominator, one can directly infer the Jacobi factor from the delta
functions and write down the deformed amplitude.

\paragraph{MHV Amplitudes.}

For MHV amplitudes, the explicit deformations can be worked out and analyzed
using~\cite{Bourjaily:2012gy} as
explained above.
Let us briefly discuss these deformations (see also~\cite{Ferro:2013dga}).

For \emph{odd multiplicities},
all $c_i$ can be expressed in terms of the $u_i$, which remain free.
The $c_i$ among themselves only satisfy the single equation
$\sum_{i=1}^nc_i=0$. When all $u$ are set to zero, no deformation remains.
Conversely, there is no deformation with all $c_i=0$.

For \emph{even multiplicities}, invariance requires that the sums of even/odd
$c_i$ vanish separately:
\begin{equation}
\sum_{\substack{i=1\\i\text{ odd}}}^nc_i=0\,,\qquad
\sum_{\substack{i=2\\i\text{ even}}}^nc_i=0\,.
\label{eq:csep}
\end{equation}
These cases admit a one-parameter family of solutions where all $c_i$
vanish, namely
\begin{equation}
\brk{u_1,\dots,u_n}=\brk{z,-z,\dots,z,-z}\,,
\label{eq:onlyu}
\end{equation}
in this case the deformed amplitudes take the form
\begin{equation}
\amp_{n}^{\text{MHV}}(z)
=
\amp_n^{\text{MHV}}\lrbrk{\frac{\spaa{12}\spaa{34}\dots\spaa{n-1\,n}}{\spaa{23}\spaa{45}\dots\spaa{n1}}}^z\,.
\end{equation}
Note that cyclic shifts of these deformations amount to flipping the
sign of $z$,
such that the deformed amplitude is invariant under two-site cyclic
shifts.

When $n=(6\bmod{4})$, still all $c_i$ can be expressed in terms of
the $u_i$, which remain unconstrained; and there is no deformation where all
$u_i$ vanish.
However when $n=(4\bmod{4})$, then not only do the $c_i$ need to satisfy
\eqref{eq:csep}, but also the even and odd $u_i$ need to satisfy
separate equations:
\begin{equation}
\sum_{\substack{i=1\\i\text{ odd}}}^n(-1)^{(i-1)/2}\,u_i=0\,,\qquad
\sum_{\substack{i=2\\i\text{ even}}}^n(-1)^{i/2}\,u_i=0\,.
\label{eq:usep}
\end{equation}
As a consequence, not all $c_i$ can be expressed in terms of
$u_i$, and in addition to the solutions \eqref{eq:onlyu}, there is a
two-parameter family of solutions with all $u_i=0$, namely
\begin{equation}
\brk{c_1,\dots,c_n}=
\brk{c\indup{o},c\indup{e},-c\indup{o},-c\indup{e},\dots,c\indup{o},c\indup{e},-c\indup{o},-c\indup{e}}\,.
\label{eq:onlyc}
\end{equation}
These deformations take the form
\begin{equation}
\amp_n^{\text{MHV}}(c^+,c^-)
=
\amp_n^{\text{MHV}}
\lrbrk{\frac{\spaa{12}\spaa{56}\dots\spaa{n\mathord{-}3,n\mathord{-}2}}{\spaa{34}\spaa{78}\dots\spaa{n\mathord{-}1,n}}}^{c^+}
\lrbrk{\frac{\spaa{45}\spaa{89}\dots\spaa{n1}}{\spaa{23}\spaa{67}\dots\spaa{n\mathord{-}2,n\mathord{-}1}}}^{c^-}\,,
\end{equation}
where $c^{\pm}\equiv (c\indup{o}\pm c\indup{e})/2$. Note that this implies that
MHV-amplitudes with $n=(4\bmod{4})$ allow for a deformation without deforming
the Yangian generators.

\paragraph{General Cells.}

The different types of deformations for MHV amplitudes discussed
above can be understood as follows: For a given $n$-point diagram with
associated permutation $\sigma$, let $P_{\sigma}$ be the finest
partition of $\brc{1,\dots,n}$ such that $\sigma$ only permutes labels
within individual parts of the partition.
Summing up the invariance conditions \eqref{eq:invcond}
for each part of the partition
results in the constraints
\begin{equation}
0=\sum_{j\in p}c_j
\qquad
\text{for all }
p\in P_\sigma\,,
\end{equation}
where $p$ denotes any single part of
$P_\sigma$. For parts with an even number of elements one finds the
further conditions
\begin{equation}
0=\sum_{j\in p}(-1)^{i_p(j)}u_j
\qquad
\text{for all }p\in P_{\sigma}\text{ with }\abs{p}\text{ even}\,,
\end{equation}
where $i_p(j)$ denotes the position of $j$ in $p$, i.e.\ the sign
alternates. For $p$ with $\abs{p}$ odd, one cannot sum the invariance
relations in a way that all $c_j$ drop out.

For MHV amplitudes, the permutation is a shift by two sites.
When $n$ is odd, the corresponding partition is
trivial, and there are no non-trivial constraints solely among $u$'s or
$c$'s. When $n$ is even, the partition simply consists of a part
that contains only odd sites and a part that contains only even
sites; hence the $c_i$ will satisfy the relations
\eqref{eq:csep}. The relations \eqref{eq:usep} solely among $u$'s will
only hold when $n=(4\bmod{4})$, in which case each of the two parts contains an
even number of elements. This kind of analysis straightforwardly
generalizes to all top-cell diagrams.

The most extreme examples of partitions arise when the permutation
satisfies $\sigma(i)=j\Leftrightarrow\sigma(j)=i$. In this case, each
part of the partition has only two elements, and the invariance
conditions simply become $c_i=-c_j$, $u_i=u_j$ for each $p=\brc{i,j}$. The
simplest example of this type is given by the four-point amplitude.
As will become clear in
\secref{sec:glue}, this case is relevant for ABJM theory~\cite{Huang:2013owa,Huang:2014xza}.

\paragraph{Remark.}

Curiously, one type of deformation was already mentioned in the
literature long before the idea of general invariance-preserving
deformations was proposed in~\cite{Ferro:2012xw,Ferro:2013dga}:
Section~6 of~\cite{Drummond:2010uq} discusses the possibility of
deformations with non-vanishing central charges $c_i$. In particular,
a specific deformation of the six-point
NMHV top cell was identified, which has alternating exponents
on the minors in the Gra{\ss}mannian integral. This specific
deformation is not compatible with a conventional BCFW decomposition
though.

\paragraph{Six-Point NMHV Integral.}

In the NMHV case ($k=3$), the momentum-twistor-space
Gra{\ss}mannian is $\grp{G}(1,n)$. For $n=6$, after gauge fixing the
$\grp{GL}(1)$ and using the bosonic delta functions to localize four
of the integrations, one obtains a one-dimensional integral.
The deformed momentum-twistor integral then takes
the form
\begin{equation}
\widehat\grass_{6,1}(\cW)
=
\int\dd c
\frac{
\delta^{4}(\chi_1+\sum_{i=2}^{5}a^*_i(c-c^*_i)\chi_i+c\chi_6)
}{
(c-c^*_2)^{1+b_2}(c-c^*_3)^{1+b_3}(c-c^*_4)^{1+b_4}(c-c^*_5)^{1+b_5}c^{1+b_6}\prod_{i=2}^5(a^*_i)^{1+b_i}
}\,,
\end{equation}
where $\cW_i=(W_i,\chi_i)$ and
\begin{align}
\nonumber c^*_2&=-\frac{\spaa{3451}}{\spaa{3456}},
&
\;c^*_3&=-\frac{\spaa{4512}}{\spaa{4562}},
&
\;c^*_4&=-\frac{\spaa{5123}}{\spaa{5623}},
&
\;c^*_5&=-\frac{\spaa{1234}}{\spaa{6234}},
\\
a^*_2&=+\frac{\spaa{3456}}{\spaa{2345}},\;
&
a^*_3&=+\frac{\spaa{4562}}{\spaa{2345}},\;
&
a^*_4&=+\frac{\spaa{5623}}{\spaa{2345}},\;
&
a^*_5&=+\frac{\spaa{6234}}{\spaa{2345}}\,,
\end{align}
with $\spaa{1234} \equiv \epsilon_{ABCD}\,W_{1}^{A}W_{2}^{B}W_{3}^{C}W_{4}^{D}$.

\section{Integrable Deformations in ABJM Theory}
\label{sec:ABJM}

In this section we initiate the investigation of integrable
deformations of scattering amplitudes in ABJM theory. Leading
singularities of the $\superN=6$ superconformal Chern--Simons
matter theory, also known as ABJM theory, are invariant under the
undeformed $\alg{osp}(6|4)$ Yangian algebra. These leading
singularities are equivalent to the residues of an integral formula:
An integral over the space of $k$ null planes in a $2k$-dimensional
space, whose integration contour localizes on the zeros of
consecutive minors of the respective $(k\times 2k)$ matrix.%
\footnote{Conformal
three-dimensional Chern--Simons matter theories have non-trivial
S-matrix elements for even multiplicity only. This is because only the
matter fields carry physical degrees of freedom and dimensional
analysis forbids cubic couplings among the matter fields.}
Tree-level
amplitudes again are given by linear combinations of these leading
singularities in which all unphysical poles cancel.

At four points, there is only one leading singularity and thus, without loss of
generality, we will consider the most general possible deformation of the
four-point amplitude that is consistent with the level-zero generators, i.e.\
the $\alg{osp}(6|4)$ superconformal symmetry.
The deformation will generically break the invariance under the original level-one generators.
However, symmetry is preserved if we deform the level-one generators
appropriately, that is if we use the evaluation representation of the Yangian
with non-vanishing evaluation parameters.

The resulting four-point deformation will serve as a template from
which we construct a deformation of the orthogonal Gra{\ss}mannian
integral and the corresponding deformed symmetry generators. It will
also serve as the fundamental building block for constructing more
general deformed Yangian invariants, which are deformations of the
residues of the undeformed Gra{\ss}mannian integral.

\subsection{Deformed Four-Point Amplitude}
\label{sec:4ptabjm}

Super-Poincar\'e invariance requires the four-point amplitude to be
proportional to the (super)momentum conserving delta functions.
Dilatation invariance constrains the proportionality function to be a
degree $-2$ polynomial of $\vev{ij}$.%
\footnote{Here,
$\vev{ij}\equiv\varepsilon_{\alpha\beta}\lambda_i^\alpha\lambda_j^\beta$,
where the spinors $\lambda_i$ parametrize the
three-dimensional massless momenta as
\mbox{$p_i^\mu=\sigma^\mu_{\alpha\beta}\lambda_i^\alpha\lambda_i^\beta$} for
symmetric matrices $\sigma^\mu$.}
We thus make the following natural ansatz for the deformed four-point amplitude:%
\footnote{There are two superfields in ABJM theory that take the
form~\cite{Bargheer:2010hn}
\begin{align}
\Phi(\Lambda)
&=
\phi^4(\lambda)
+\eta^A\psi_A(\lambda)
+\half\eps_{ABC}\eta^A\eta^B\phi^C(\lambda)
+\sfrac{1}{6}\eps_{ABC}\eta^A\eta^B\eta^C\psi_4(\lambda)\,,
\\
\bar\Phi(\Lambda)
&=
\bar\psi^4(\lambda)
+\eta^A\bar\phi_A(\lambda)
+\half\eps_{ABC}\eta^A\eta^B\bar\psi^C(\lambda)
+\sfrac{1}{6}\eps_{ABC}\eta^A\eta^B\eta^C\bar\phi_4(\lambda)\,.
\end{align}
Here $\Phi$ is bosonic while $\bar\Phi$ is fermionic.}
\begin{align}
\label{eq:4ptMom}
\amp_4(\bar\Phi_1,\Phi_2,\bar\Phi_3,\Phi_4)(\cc)
&=
\frac{\deltaPQ}{\spaa{12}^{1+\cc}\spaa{23}^{1-\cc}}
\equiv
\deltaPQ\,f(\lambda).
\end{align}
Following~\cite{Bargheer:2010hn}, it is straightforward to see that this is invariant under the
superconformal boost (level-zero) generators, and thus under the full
$\alg{osp}(6|4)$ level-zero symmetry algebra.

\paragraph{Invariance under the Level-One Momentum Generator.}

For Yangian invariance, we only need to show the invariance under the level-one momentum generator;
invariance under all other level-one generators then follows from the
commutation relations and the level-zero invariance.
The level-one momentum generator in the evaluation representation
takes the form~\cite{Bargheer:2010hn}
\begin{equation}
\label{eq:P1}
\gen{\widehat P}^{\alpha\beta}
=\sum_{1\leq j<k\leq n}\half
\lrsbrk{
\bigbrk{\gen{L}^{(\alpha}_{j\gamma}+\delta^{(\alpha}_\gamma\gen{D}_j}\gen{P}_k^{\gamma\beta)}
-\gen{Q}_j^{(\alpha A}\gen{Q}_k^{\beta)}{}_A
-(j\leftrightarrow k)
}
+\sum_k u_k \gen{P}_k^{\alpha\beta}\,.
\end{equation}
The single-site generators are given in \appref{sec:ABJMGen}. Using the
transformation properties of the delta functions, the symmetry equation
simplifies to%
\footnote{Here we use the notation
$X^{(\alpha\beta)}=X^{\alpha\beta}+X^{\beta\alpha}$ and
$X^{[\alpha\beta]}=X^{\alpha\beta}-X^{\beta\alpha}$.}
\begin{equation}
\gen{\widehat P}^{\alpha\beta}\amp_4(\cc)
=
\deltaPQ\biggbrk{
\sum_{1\leq j<k\leq 4}
\Bigsbrk{
\half\gen{P}_k^{\gamma(\beta}\bigbrk{\lambda_j^{\alpha)}\partial_{j\gamma}
+\half\delta_\gamma^{\alpha)}}f(\lambda)
-(j\leftrightarrow k)
}
+\sum_{k=1}^4 u_k\gen{P}_k^{\alpha\beta} f(\lambda)
}.
\label{eq:symeq4pt}
\end{equation}
We know from~\cite{Bargheer:2010hn} that the undeformed amplitude is
invariant under $\gen{\widehat P}|_{u_k=0}$, and thus only the terms
proportional to $\cc$ remain when acting with $\gen{\widehat
P}|_{u_k=0}$ on the deformed amplitude. These terms are generated when
the bosonic derivatives act on the denominator of $\amp_4(\cc)$. We
can collect the terms from the first term in the square bracket as follows:
\begin{equation}
U_k^{\alpha\delta}
\equiv\sum_{j=1}^{k-1}\lambda_j^\alpha\partial_j^\delta f(\lambda)
=
\begin{cases}
0+\dots\,, & k=1\,,\\
-\cc\frac{\lambda_1^\alpha\lambda_2^\delta}{\spaa{12}}f(\lambda)+\dots, & k=2\,,\\
\Bigbrk{-\cc\frac{\lambda_1^{[\alpha}\lambda_2^{\delta]}}{\spaa{12}}+\cc\frac{\lambda_2^\alpha\lambda_3^\delta}{\spaa{23}}}f(\lambda)+\dots, & k=3\,,\\
\Bigbrk{-\cc\frac{\lambda_1^{[\alpha}\lambda_2^{\delta]}}{\spaa{12}}+\cc
\frac{\lambda_2^{[\alpha}\lambda_3^{\delta]}}{\spaa{23}}}f(\lambda)+\dots, & k=4\,.
\end{cases}
\end{equation}
Here we only display the terms proportional to $\cc$ since the rest is known to
cancel in the undeformed limit.
Evaluating the symmetric and anti-symmetric contributions from
$U_k^{\alpha\delta}$ separately, we find%
\footnote{We use that $\eps_{\gamma\delta}\eps^{\alpha\delta}=-\delta_\gamma^\alpha$
and $\lambda_j^{[\alpha}\lambda_k^{\beta]}=-\eps^{\alpha\beta}\spaa{jk}$,
where we define $\eps_{12}=1=-\eps^{12}$.}
\begin{gather}
\half\sum_{k=1}^4\gen{P}_k^{\gamma\beta}\eps_{\gamma\delta}U_k^{(\alpha\delta)}+(\alpha\leftrightarrow\beta)
=+\cc\bigsbrk{\gen{P}_2^{\alpha\beta}-\gen{P}_3^{\alpha\beta}}f(\lambda)\,,
\nn\\
\half\sum_{k=1}^4\gen{P}_k^{\gamma\beta}\eps_{\gamma\delta}U_k^{[\alpha\delta]}+(\alpha\leftrightarrow\beta)
=-\cc\bigsbrk{\gen{P}_2^{\alpha\beta}+\gen{P}_3^{\alpha\beta}}f(\lambda)\,.
\label{eq:U}
\end{gather}
Repeating the analysis for the term with $(j\leftrightarrow k)$ in
\eqref{eq:symeq4pt}, we find (we relabel the summation indices)
\begin{equation}
\bar U_k^{\alpha\delta}
\equiv\sum_{j=k+1}^{4}\lambda_j^\alpha\partial_j^\delta f(\lambda)
=
\begin{cases}
\cc\Bigbrk{\frac{\lambda_2^{\alpha}\lambda_1^{\delta}}{\spaa{12}}+\frac{\lambda_2^{[\alpha}\lambda_3^{\delta]}}{\spaa{23}}}f(\lambda)+\dots, & k=1\,,\\
-\cc\frac{\lambda_3^{\alpha}\lambda_2^{\delta}}{\spaa{23}}f(\lambda)+\dots, & k=2\,,\\
0+\dots\,, & k=3,4\,,
\end{cases}
\end{equation}
and thus
\begin{gather}
-\half\sum_{k=1}^4\gen{P}_k^{\gamma\beta}\eps_{\gamma\delta}\bar U_k^{(\alpha\delta)}+(\alpha\leftrightarrow\beta)
=+\cc\bigsbrk{-\gen{P}_1^{\alpha\beta}+\gen{P}_2^{\alpha\beta}}f(\lambda)\,,
\nn\\
-\half\sum_{k=1}^4\gen{P}_k^{\gamma\beta}\eps_{\gamma\delta}\bar U_k^{[\alpha\delta]}+(\alpha\leftrightarrow\beta)
=-\cc\bigsbrk{+\gen{P}_1^{\alpha\beta}+\gen{P}_2^{\alpha\beta}}f(\lambda)\,.
\label{eq:Ubar}
\end{gather}
Combining the results from \eqref{eq:U,eq:Ubar}, we finally arrive at
\begin{align}
\gen{\widehat P}^{\alpha\beta}\amp_4(\cc)
=\deltaPQ
\Bigsbrk{&
\gen{P}_1^{\alpha\beta}(u_1-\cc)
+\gen{P}_2^{\alpha\beta}u_2
+\gen{P}_3^{\alpha\beta}(u_3-\cc)
+\gen{P}_4^{\alpha\beta}u_4
}f(\lambda)\,.
\end{align}
Hence, requiring this expression to be proportional to the total momentum
acting on the deformed amplitude (i.e.\ to vanish), we find the following
constraints on the parameters following from invariance under the level-one
momentum generator:
\begin{align}
u_1-\cc\,=
u_2\,=
u_3-\cc\,=
u_4\,=\text{const}\,.
\end{align}
Alternatively, these constraints can be expressed as
\begin{equation}
\eqbox{u_k-u_{k-1}=(-1)^{k-1}\cc\,,
\qquad
k=1,\dots,4\,.}
\label{eq:4ptsol}
\end{equation}
In conclusion, the deformed four-point amplitude in \eqref{eq:4ptMom} is
invariant under the evaluation representation of the Yangian generators of
$Y[\alg{osp}(6|4)]$, provided that the level-one generators are deformed as in
\eqref{eq:P1} with the parameters $u_i$ related to $z'$ via \eqref{eq:4ptsol}.

Note that the deformation \eqref{eq:4ptMom} changes the weight of
$\amp_4(\bar\Phi,\Phi,\bar\Phi,\Phi)$ under
$\exp\brk{i\pi\lambda_i\cdot\partial/\partial\lambda_i}$ for $i=1,3$, but not for
$i=2,4$. That is, the deformation deforms the phase of the fermionic
legs, but preserves the phase of the bosonic legs. For further
comments on this, see the discussion around \eqref{eq:scalingop} below.

In the next two sections we will discuss deformed invariants at higher
multiplicities. First we will construct bigger deformed on-shell
diagrams by gluing four-point vertices. Next we show the invariance of
the deformed Gra{\ss}mannian integral explicitly.

\subsection{Gluing Invariants}
\label{sec:glue}

All ABJM on-shell diagrams can be constructed by iteratively gluing
four-point vertices together~\cite{ArkaniHamed:2012nw,Huang:2013owa}.
Along the lines of the four-dimensional case~\cite{Beisert:2014qba}
reviewed in \secref{sec:defon4d} above,
the gluing procedure can be split into two steps that need to be
iterated: Taking products of diagrams, and fusing lines. In the
following we will show that the gluing procedure indeed preserves the
Yangian invariance also in the deformed case, provided that the
deformation parameters are identified appropriately. For showing
invariance, we will use the completely general form \eqref{eq:lev1_J}
of the $n$-point Yangian level-one generators.

\paragraph{Products.}

Given two diagrams $\cY_1(1,\dots,m)$ and
$\cY_2(m+1,\dots,n)$ that are invariant under the $m$-point
and $(n-m)$-point Yangian algebras with evaluation
parameters $\brc{u_1,\dots,u_m}$ and $\brc{u_{m+1},\dots,u_n}$, the
product
\begin{equation}
\cY'(1,\dots,n)
=
\cY_1(1,\dots,m)
\,\cY_2(m+1,\dots,n)
\end{equation}
is invariant under the $n$-point Yangian algebra with
evaluation parameters $\brc{u_1,\dots,u_n}$:
\begin{equation}
\gen{\widehat J}^a\cY'
=
\bigbrk{\gen{\widehat J}^a\cY_1}\cY_2
+\cY_1\bigbrk{\gen{\widehat J}^a\cY_2}
+f^a{}_{bc}\bigbrk{\gen{J}^a\cY_1}\bigbrk{\gen{J}^b\cY_2}
=0\,.
\end{equation}
%

\paragraph{Fusion.}

From any invariant $(n+2)$-point diagram
$\cY(1,\dots,n,n+1,n+2)$, one can construct an
$n$-point diagram by fusing two adjacent external lines,
\begin{equation}
\cY'(1,\dots,n)
=
\int\dL\,\dL'
\,\delta^{2|3}(\Lambda-i\Lambda')
\,\cY(1,\dots,n,\Lambda,\Lambda')\,.
\end{equation}
Here we will use the kinematical variables
$\Lambda^{\cA}=(\lambda^{\alpha},\eta^A)$
with $\alpha=1,2$ and $A=1,2,3$.%
\footnote{Some care needs to be taken in the definition of the
on-shell integration over $\dd^{2|3}\Lambda$, see e.g.~\cite{Bargheer:2012cp}.
Throughout this work, such integrations will
always be localized on delta functions.}
The diagram $\cY'$ will be invariant under the $n$-point
Yangian algebra with evaluation parameters $\brc{u_1,\dots,u_n}$
provided that
\begin{equation}
\eqbox{u_{n+1}=u_{n+2}\,.}
\label{eq:gluecond}
\end{equation}
Both the level-zero and the level-one invariance of
$\cY'$ can be shown
straightforwardly, using the $(n+2)$-point invariance of
$\cY$ and the fact that
\begin{equation}
\int\dL\,\dL'
\,\delta^{2|3}(\Lambda-i\Lambda')
\,\bigbrk{\gen{J}_{\Lambda}^a+\gen{J}_{\Lambda'}^a}
\,f(\Lambda,\Lambda')
=0
\label{eq:pi}
\end{equation}
for any function $f$.
The latter can be verified directly with the explicit
$\alg{osp}(6|4)$ generators given in \appref{sec:ABJMGen}.
Using the invariance of $\cY$, the action of the level-one
generator on $\cY'$ can be written as
\begin{multline}
\gen{\widehat J}^a_{1\dots n}\,\cY'(1,\dots,n)
=
-f^a{}_{bc}\int\dL\,\dL'
\,\delta^{2|3}(\Lambda-i\Lambda')
\\\cdot
\lrbrk{\sum_{i=1}^n\gen{J}_i^b\bigbrk{\gen{J}_{\Lambda}^c+\gen{J}_{\Lambda'}^c}+\gen{J}_{\Lambda}^b\gen{J}_{\Lambda'}^c}
\cY(1,\dots,n,\Lambda,\Lambda')\,.
\label{eq:fusion1}
\end{multline}
The first term in the parentheses vanishes due to \eqref{eq:pi}. Again
using \eqref{eq:pi}, the second term can be rewritten as
$\gen{J}_{\Lambda}^b\gen{J}_{\Lambda}^c\simeq\half\comm{\gen{J}_\Lambda^b}{\gen{J}_\Lambda^c}=\half
f^{bc}{}_d\gen{J}^d$. Hence,
this term is proportional to $f^a{}_{bc}f^{bc}{}_{d}$, which vanishes
for $\alg{osp}(6|4)$, as it does for $\alg{psu}(2,2|4)$, since the dual
Coxeter number is zero.

The above procedure of taking products and fusing lines allows us to fuse
two legs from different diagrams, or two legs sitting on the same
diagram. Note, however, that the two fused legs have to correspond to
different multiplets, i.e.\ to fields $\Phi$ and $\bar\Phi$, and that
they must be adjacent.

\paragraph{Four-Vertex.}

We have seen above that the deformed Yangian-invariant four-vertex
reads%
\footnote{The last equality follows from $\spaa{ij}=\pm\spaa{kl}$,
$\spaa{jk}=\pm\spaa{li}$ (with aligned signs) due to momentum
conservation.}
\begin{equation}
\amp_4(\bar\Phi_i,\Phi_j,\bar\Phi_k,\Phi_\ell)(z')
=
\includegraphicsbox{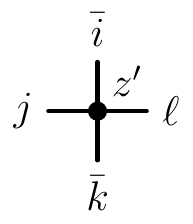}
=
\frac{\delta^3(P)\,\delta^6(Q)}{\spaa{ij}^{1+z'}\spaa{jk}^{1-z'}}
=
\frac{\delta^3(P)\,\delta^6(Q)}{\spaa{kl}^{1+z'}\spaa{li}^{1-z'}}\,,
\label{eq:4vert}
\end{equation}
where the Yangian evaluation parameters $u_{i,j,k,\ell}$ need to satisfy
\begin{equation}
u_i=u_k\,,
\qquad
u_j=u_\ell\,,
\qquad
z'=u_i-u_j\,.
\end{equation}
All bigger deformed on-shell diagrams can be constructed from this
four-vertex by applying the invariance-preserving operations described
above. For this purpose, it is most useful to write the vertex
\eqref{eq:4vert} in a gauge-fixed integral form%
\footnote{The domain of integration has to be chosen such that the
delta functions localize the integral on a single point. For real
kinematics in Minkowski space, a valid choice for the integration
domain is $[0,\pi)$.}
\begin{equation}
\amp_4(z)=\int\frac{\dd\theta}{\sin(\theta)^{1+z}}\,\delta^{4|6}\bigbrk{C(\theta)\cdot\Lambda}\,,
\label{eq:4vertint}
\end{equation}
where the $C$-matrix $C(\theta)$ is given by~\cite{Huang:2013owa}
\begin{equation}
\label{eq:Cmatrix_4pt}
C(\theta)=
\begin{pmatrix}
1 & 0 & i\sin \theta & i\cos\theta \\
0 & 1 & -i\cos\theta & i\sin\theta
\end{pmatrix}
\,.
\end{equation}
It is easy to see that $z=-z'$ for this choice of $C$-matrix.
In general, the relative sign
between $z$ and $z'$ depends on which columns of $C(\theta)$ are set to unit
vectors. Therefore, as we build up a general on-shell diagram, we need
to keep track of which columns of $C(\theta)$ are set to unity. A
convenient way of keeping track is to decorate the lines connected to
each vertex with two incoming and two outgoing arrows, where the
former indicates that these columns form the unit matrix. We will only
consider the cases where the two incoming arrows are adjacent, which
leads to a constraint on $z$ included in the following figure:
\begin{equation}
\includegraphicsbox{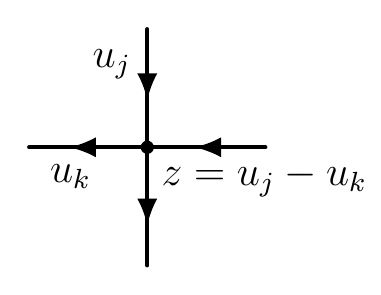}.
\label{eq:fig4vertex}
\end{equation}
The sign of $z$ is determined by the following rule: Start with a line
associated with the parameter $u_j$ and compare its arrow to the clockwise
neighboring line associated with $u_k$. If both arrows have the same
orientation with respect to the vertex, then $z=u_k-u_j$; if the
arrows have opposite orientations, then $z=u_j-u_k$.
As we will see further below, the lines in \eqref{eq:fig4vertex}
will be identified with the rapidity lines of integrable models,
and the parameters $u_i$ with rapidity parameters.

\paragraph{General Deformed Diagrams.}

\begin{figure}[t]\centering
\includegraphicsbox{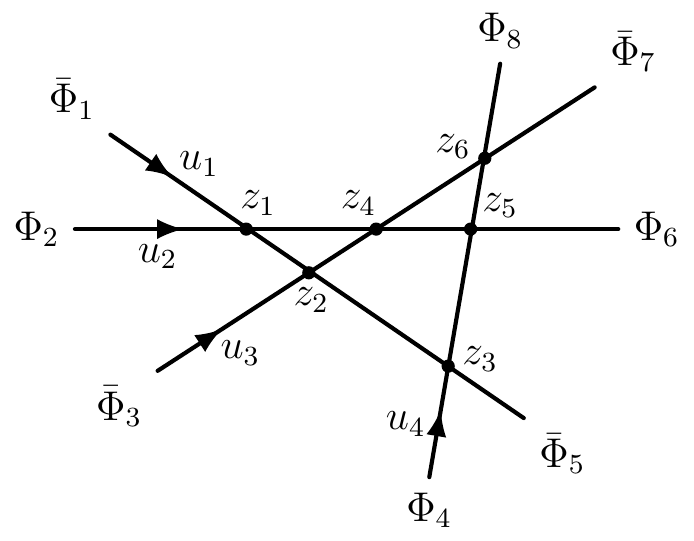}\qquad
\caption{An example of a deformed on-shell diagram with the
invariance constraints given by \protect\eqref{eq:10usol}.}
\label{fig:exdia}
\end{figure}

Any reduced $2k$-point on-shell diagram of ABJM theory can be drawn as
$k$ straight lines that intersect,%
\footnote{For reduced diagrams, any two lines intersect at most once.}
where each intersection is a
four-point on-shell vertex.
Turning on the deformations, there is one deformation modulus $z_i$
for each four-vertex, where~$i$ labels the vertices in the respective
diagram. For the larger diagrams, the invariance
conditions \eqref{eq:fig4vertex} for each four-vertex, and the invariance
conditions \eqref{eq:gluecond} from fusing lines must be respected. It
immediately follows that for every invariant $2k$-point diagram, there
remains exactly one evaluation
parameter for each of the $k$ straight lines, as the evaluation
parameters $u_i$ on glued lines need to be identified.
For example for the following diagram we have:
\begin{center}
\includegraphicsbox{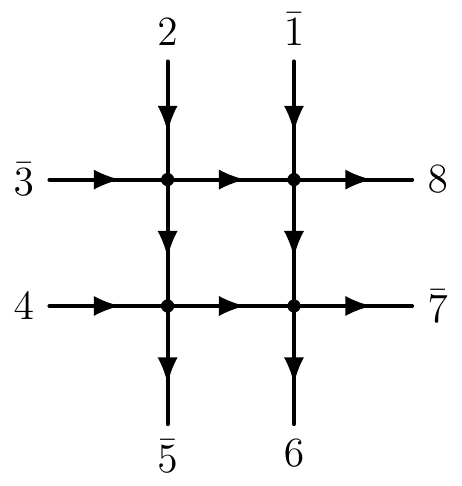}
\qquad
\autoparbox{
\begin{minipage}{4cm}
\begin{align*}
u_{1}&=u_{6}\,,\\
u_{2}&=u_{5}\,,\\
u_{3}&=u_{8}\,,\\
u_{4}&=u_{7}\,.\\
\end{align*}
\end{minipage}
}
\end{center}
Each vertex deformation
modulus $z_i$ is in turn determined to be the difference of the
evaluation parameters on the lines that pass through the vertex according to \eqref{eq:fig4vertex}.
As an example, the $z_i$'s in \figref{fig:exdia} are given by
\begin{align}
\label{eq:10usol}
\nonumber
z_1&=u_2-u_1\,,\quad z_2=u_3-u_1\,,\quad z_3=u_4-u_1\,,\\
z_4&=u_3-u_2\,,\quad z_5=u_4-u_2\,,\quad z_6=u_4-u_3\,.
\end{align}
For generic diagrams,
the conditions \eqref{eq:fig4vertex} not only determine the vertex
moduli, but also induce constraints among them: For each closed loop,
$\sum_i(\pm z_i)=0$, where $i$
enumerates the
vertices along the loop, and where the sign depends on the relative
directions of arrows along the loop at the respective vertex.

In summary, every $2k$-point diagram admits a $(k-1)$-parameter family
of deformations, where the $(k-1)$ parameters are given by the
evaluation parameters $u_{1\dots k}$ on the $k$ lines, modulo a trivial overall
shift of all $u_i$'s.

As described in~\cite{Huang:2013owa}, the vertex
\eqref{eq:4vertint} provides ``canonical
coordinates,'' which means that gluing multiple such vertices
produces no Jacobian from combining the delta
functions; that is a general (deformed) diagram constructed in this
way takes the simple form
\begin{equation}
\cY(\Lambda)
=
\int
\frac{\dd\theta_1}{\sin(\theta_1)^{1+z_1}}
\dots
\frac{\dd\theta_\ell}{\sin(\theta_\ell)^{1+z_\ell}}
\delta^{2k|3k}\bigbrk{C(\theta_i)\cdot\Lambda}\,,
\label{eq:diagint}
\end{equation}
where the orthogonal matrix $C(\theta_i)$ can be read off algorithmically from the diagram.

\paragraph{Deformed Triangle Move.}

Undeformed on-shell diagrams are invariant under triangle moves, which
take one line past the intersection
of two other lines~\cite{ArkaniHamed:2012nw,Huang:2013owa}.
The triangle move amounts to
a change of integration variables in the Gra{\ss}mannian integral that
preserves the canonical form \eqref{eq:diagint} in the undeformed
case. Not surprisingly, this remains true without modifications in the deformed case:
\begin{equation}
\includegraphicsbox{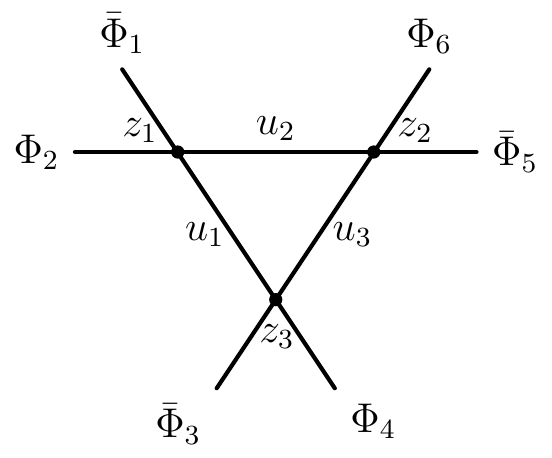}
\quad\text{\Large $=$}\quad
\includegraphicsbox{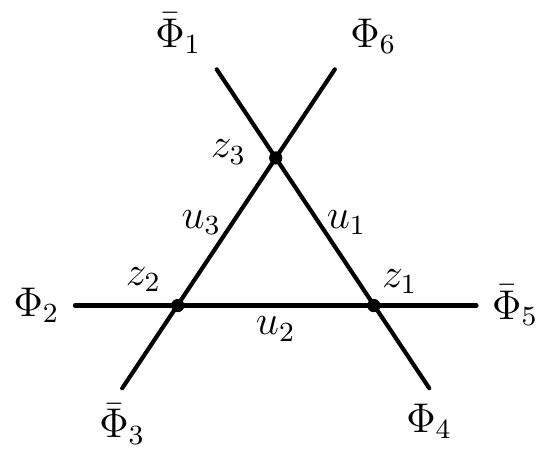}\,.
\label{eq:trianglemove}
\end{equation}
The triangle equality holds regardless of the orientations of the
three lines, as long as the orientations are the same on both sides of
the equation.
Note that this result is consistent with the invariance constraints
\eqref{eq:fig4vertex}. In fact, it is not a coincidence that this
diagrammatic equation looks very much like the Yang--Baxter equation,
as will become more clear in \secref{sec:RmatABJM} below.

\paragraph{Invariance and Permutations.}

Due to the triangle equality \eqref{eq:trianglemove}, every (deformed)
reduced diagram is uniquely specified by a permutation
$\sigma$ that simply interchanges pairs of external legs. In other
words, $\sigma$ is composed of pairwise commuting transpositions, and
$\sigma^2$ equals the identity permutation.
The invariance equations for the evaluation parameters then take the
rather trivial form
\begin{equation}
\label{eq:u_perm}
\eqbox{u_i=u_{\sigma(i)}\,.}
\end{equation}
In order to identify the vertex deformation moduli $z_i$, one needs to
decorate the $2k$-point diagram with arrows such that each
line (connecting legs $i$ and $\sigma(i)$)
carries a definite orientation. Then $k$ columns of the
$C$-matrix form the identity matrix, and each four-vertex is of the
form \eqref{eq:fig4vertex}, such that the $z_i$ can be read off.

\paragraph{Deformed BCFW Decomposition.}

Tree-level amplitudes in ABJM theory can be decomposed into a sum of
BCFW terms~\cite{Gang:2010gy}, where each term is an on-shell
diagram~\cite{Huang:2013owa}. An interesting question is whether higher-point
tree-level amplitudes can be consistently deformed by deforming each
diagram in the sum, using the same evaluation parameters for each
diagram. The six-point amplitude consists of a single triangle-shaped
diagram as in \eqref{eq:trianglemove}, and thus allows for a two-parameter family of deformations.
The diagrams for the eight- and ten-point amplitudes are given
explicitly in~\cite{Huang:2013owa}.
The eight-point tree-level amplitude consists of two terms, in which
the lines connect the eight points as $\brc{[15][27][36][48]}$,
$\brc{[14][26][37][58]}$. Hence it allows for a one-parameter
deformation in terms of $u_1-u_2$, where $u_1=u_4=u_5=u_8$, and
$u_2=u_3=u_6=u_7$. The ten-point amplitude consists of the five terms:
\begin{align}
\nonumber
&\brc{[14][27][39][58][6,10]},\;\;\brc{[14][26][38][59][7,10]},\;\;
\brc{[16][29][37][4,10][58]},\\
&\qquad\qquad\qquad\quad\brc{[17][29][36][48][5,10]},\;\;
\brc{[15][28][36][49][7,10]}.
\end{align}
Combining the resulting invariance constraints enforces
that all evaluation parameters must be equal, and hence there is no non-trivial
deformation at ten points.
In fact, the $(2p+4)$-point tree-level
amplitude consists of $(2p)!/(p!(p+1)!)$ diagrams~\cite{Huang:2013owa}. Since each diagram
implies a different set of constraints, the number of constraints at
higher points by
far outweighs the number of parameters, and hence a
consistent deformation of the BCFW decomposition beyond eight
points cannot be expected.

\paragraph{Branches.}

Similar to the $\superN=4$ SYM case, every ABJM on-shell diagram is an
integral over a cell in the orthogonal
Gra{\ss}mannian~\cite{ArkaniHamed:2012nw,Huang:2013owa}.
Every cell in the orthogonal Gra{\ss}mannian in fact consists of two
distinct branches. The two branches can be distinguished by the ratios
of non-overlapping minors
\begin{equation}
M_j/M_{j+k}=\mp 1\,.
\end{equation}
In gluing on-shell diagrams, this subtlety is reflected in the matrix
$C(\theta)$ of the four-point vertex \eqref{eq:4vertint}, where
\begin{equation}
\label{eq:OG2}
\grp{OG}_{2,\pm}:
\qquad
C(\theta)
=
\begin{pmatrix}
1 & 0 &\pm i\sin\theta & \pm i\cos\theta\\
0 & 1 & -i\cos\theta & i\sin\theta
\end{pmatrix}
\,.
\end{equation}
Note that these two $C$-matrices are not related by any coordinate
transformation. While it may appear that there are $2^{n_\text{v}}$
distinct branches for a given on-shell diagram with $n\indup{v}$
vertices, most of them are related by coordinate transformations,
leaving only two distinct branches. Denoting the branches at each
vertex by a sign, the branch of the final $C$-matrix is simply the
product of all signs of the individual vertices. Since each branch is
individually Yangian invariant, we restrict ourselves to
diagrams built from the positive branch of the four-vertex.
Generalizing to include the other branch is straightforward.

\subsection{Deformed Orthogonal Gra{\ss}mannian}
\label{sec:defgrass3d}

In this section we consider integrable deformations of the orthogonal
Gra{\ss}mannian integral of ABJM theory. As in four dimensions, the
deformation under consideration is again a modification of the power
of the minors, which are the only admissible deformations that
maintain $\grp{GL}(k)$ invariance. We will first map the deformation
parameters of the four-point Gra{\ss}mannian to that of the four-point
amplitude. Then we will show that for general multiplicities, the
deformed Gra{\ss}mannian is invariant under Yangian symmetry, provided
that the deformation parameters obey a set of constraints that are a
generalization of the four-point constraints~\eqref{eq:4ptsol}.

\paragraph{Proposal for the Deformed Gra{\ss}mannian.}

We consider the following deformation of the orthogonal Gra{\ss}mannian integral:
\begin{align}
\label{Int}
\grass_{2k}(b_i)=\int\frac{d^{k\times 2k} C}{|\GL(k)|}
\frac{\delta^{k(k+1)/2}(C\cdot C\transpose)\,\delta^{2k|3k}(C\cdot\Lambda)}
{\prod_{i=1}^{k} M_i^{1+b_i}}\,.
\end{align}
The undeformed integral was originally proposed in~\cite{Lee:2010du}.
Here $C\cdot C\transpose\equiv\sum_{i}C_{a i}C_{b i}$ is a
$k\times k$-symmetric matrix whose vanishing implies that the
Gra{\ss}mannian matrix $C$ consists of $k$ $n$-dimensional
null vectors. We will denote the orthogonal Gra{\ss}mannian G($k,2k$)
as $\grp{OG}_k$. For the integral to be $\GL(k)$ invariant, the deformation
parameters $b_i$ must satisfy the relation $\sum_{i=1}^{k} b_i=0$.
When all $b_i$ vanish, this reduces to the formula given in~\cite{Lee:2010du}.

\paragraph{Relation to the Four-Point Amplitude.}

For the simplest case of $k=2$, the relation between $\cc$ in \eqref{eq:4ptMom} and $b_1$ in
\eqref{Int} can be deduced by simply using the bosonic delta functions
to localize the Gra{\ss}mannian integration variables. More precisely, let us
begin with the following integral, where we already used that $b_1+b_2=0$ in
$\eqref{Int}$:
\begin{equation}
\label{4ptInt}
\grass_4(b_1,-b_1)=\int\frac{d^{2\times4} C}{|\GL(2)|}
\frac{\delta^{3}(C\cdot C\transpose)\,\delta^{4|6}(C\cdot\Lambda)}
{ M_1^{1+b_1} M_2^{1-b_1}}\,.
\end{equation}
We work with the gauge
\begin{align}
C=
\begin{pmatrix}
1 & 0 & C_{13} & C_{14}\\
0 & 1 & C_{23} & C_{24}
\end{pmatrix}
,
\end{align}
and the momentum delta function
$\delta^4(C\cdot\lambda)$ gives~\cite{Gang:2010gy}
\begin{align}
\label{eq:delta4}
\delta^4(C\cdot\lambda)=\frac{1}{\spaa{34}^2}
\prod_{r,s}\delta^4\lrbrk{C_{r,s}-C_{r,s}^*},
\quad
\begin{pmatrix}
C^*_{13} & C^*_{14}\\
C^*_{23} & C^*_{24}
\end{pmatrix}
=
-\frac{1}{\spaa{34}}
\begin{pmatrix}
\spaa{14} & \spaa{31}\\
\spaa{24} & \spaa{32}
\end{pmatrix}
.
\end{align}
Substituting the solutions into \eqref{4ptInt}, we find the following deformed amplitude:
\begin{align}
\cA_4(b_1)=
\frac{\delta^3(P)\delta^6(Q)}
{\spaa{12}^{1+b_1}\spaa{23}^{1-b_1}}\,.
\end{align}
Setting $b_1=z'$,
we see that the above deformation of the Gra{\ss}mannian indeed induces the
same deformed four-point amplitude as in \eqref{eq:4ptMom}.

\paragraph{Relation to Deformed On-Shell Diagrams.}

\begin{figure}\centering
\includegraphicsbox{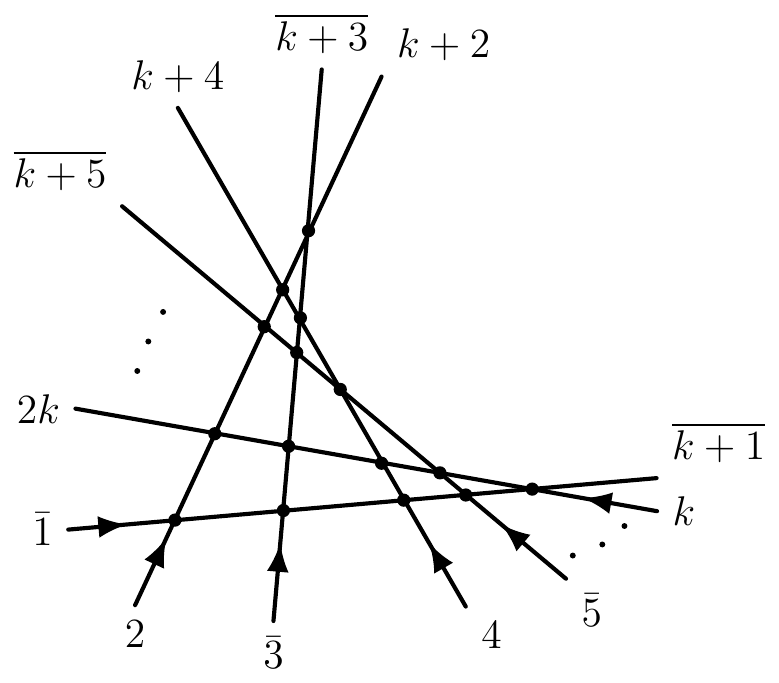}
\caption{The ABJM top-cell diagram.}
\label{fig:3dtopcell}
\end{figure}
For higher multiplicities, we first note that the deformation of the
Gra{\ss}mannian integral can be obtained from the deformed on-shell
diagrams simply by considering the top-cell diagram.
The $2k$-point top cell has dimension $k(k-1)/2$. All
$2k$-point diagrams consist of $k$ lines, and each four-vertex
contributes one integration variable. Hence, in the top-cell diagram,
each of the $k$ lines has to cross each other line exactly once.
Modulo triangle moves, this diagram is unique. A canonical
representative is sketched in \figref{fig:3dtopcell}.
Iteratively building up the top cell by gluing four-vertices
\eqref{eq:4vertint}, the top-cell integral will take the form
\eqref{eq:diagint}, with $\ell=k(k-1)/2$. Comparing that form to the
deformed Gra{\ss}mannian integral formula \eqref{Int},
one could in principle read off the relation between $b_i$ and $z_i$, and in turn
express $b_i$ in terms of the Yangian evaluation parameters $u_i$.

In four
dimensions, we saw that the invariance conditions for the top cell directly lead to the
simple relation \eqref{eq:topcellexp} between the exponents $b_i$ and
the parameters $u_i^\pm$, which are the natural deformation moduli
from the on-shell diagram perspective. The relation followed from a
direct identification of the central charges $c_i$ in terms of the
exponents $b_i$.

However, the symmetry algebra $\alg{osp}(6|4)$ of ABJM theory does not
admit a central extension, and thus there are no central charges that
could be deformed.%
\footnote{More concretely, the scattering amplitudes and the
Gra{\ss}mannian integral for ABJM are not eigenstates of the local scaling operators
$\Lambda_i\cdot\partial/\partial\Lambda_i$; see also the discussion
around \eqref{eq:scalingop} below.}
Hence, the relation between the exponents $b_i$ and the evaluation
parameters $u_i$ apparently cannot be deduced directly. Below, we will
therefore derive the invariance relations by directly acting with the
Yangian generators on the deformed Gra{\ss}mannian integral.

\paragraph{Yangian Invariance of the Deformed Gra{\ss}mannian.}

For the four-point example, we have deduced that the relation between
the deformation parameter of the Gra{\ss}mannian integral and the
evaluation parameters is given by \eqref{eq:4ptsol}, with $b_1=\cc$.
We now proceed to derive the general $n$-point relations.

For compactness we present the level-zero and level-one generators in $\Lambda$-space.
For the invariance under the level-zero algebra, note that
the level-zero generators take the form
\begin{align}
\label{eq:ABJM_level_zero}
\Lambda_i^{\cA}\Lambda_i^{\cB}\,,
\quad
\Lambda_i^{\cA}\frac{\partial}{\partial\Lambda_i^{\cB}}\,,
\quad
\frac{\partial}{\partial\Lambda_i^{\cA}}\frac{\partial}{\partial\Lambda_i^{\cB}}\,.
\end{align}
The invariance follows, respectively, from the momentum conservation,
delta-function constraint $\delta^{2k|3k}(C\cdot\Lambda)$, and the
orthogonality of $C$~\cite{Lee:2010du}.

The undeformed level-one generators $\gen{\widehat J}^{\cA\cB}$
with two upper indices of the same statistics
($\gen{P}^{ab}$ and $\gen{R}^{AB}$)
can be written as~\cite{Bargheer:2010hn}
\begin{align}
\label{J1Def1}
\gen{\widehat J}^{(\cA\cB]}
=
\Bigbrk{\sum_{l<i}-\sum_{i<l}}
\biggbrk{
(-1)^{|\cC|}
\Lambda^{(\cA}_l
\frac{\partial}{\partial\Lambda_l^{\cC}}
\Lambda_i^{\cC}
\Lambda^{\cB]}_i
+\frac{\Lambda^{(\cA}_i\Lambda^{\cB]}_i}{2}
}
\,,
\end{align}
where the indices $(\cA$, $\cB]$ are understood to be
\sbrk{anti}symmetrized if $\cA$ and $\cB$ denote indices of
$\alg{sp}(4)$ $\sbrk{\alg{su}(3)}$. Here we will simply consider the case
where they have the same statistics, since invariance under all other
generators follows from the $\alg{osp}(6|4)$ algebra.%
\footnote{The level-one generators $\gen{\widehat J}$ transform in the
adjoint representation of the level-zero algebra.}
We first rewrite $(\sum_{l<i}-\sum_{i<l})=2\sum_{l<i}-\sum_{i,l}+\sum_{i=l}$.
Then for $\gen{\widehat J}^{(\cA\cB]}$, this amounts to
\begin{equation}
\label{J1Def}
\gen{\widehat J}^{(\cA\cB]}
\simeq
\lrsbrk{
2\biggbrk{
\sum_{l<i}
\Lambda^{(\cA}_l\Lambda^{\cB]}_i\Lambda_i^{\cC}\frac{\partial}{\partial
\Lambda_l^{\cC}}+\frac{\Lambda^{(\cA}_i\Lambda^{\cB]}_i}{2}
}
+\sum_{i}\Lambda^{(\cA}_i\Lambda^{\cB]}_i\Lambda_i^{\cC}
\frac{\partial}{\partial\Lambda_i^{\cC}}
},
\end{equation}
up to terms proportional to level-zero generators.
As the invariance of the undeformed orthogonal Gra{\ss}mannian was not
proved in the literature, we provide the details in
\appref{sec:Grass}. The crucial step in the proof, as pointed
out for the four-dimensional case in~\cite{Drummond:2010qh}, is to realize that
$\Lambda_l^{\cC}\,\partial/\partial\Lambda_i^{\cC}$ acts
only on the delta functions and can be converted into a
rotation generator acting on the Gra{\ss}mannian variables:
\begin{equation}
\label{ODef}
\Lambda_i^{\cC}\frac{\partial}{\partial\Lambda_l^{\cC}}
\quad\longrightarrow\quad
O_{l}\,^{i}\equiv\sum_{a}C_{al}\frac{\partial}{\partial C_{ai}}\,.
\end{equation}
Using integration by parts, the linear operator then acts on the
integration measure. This operator simply replaces column $i$ of the
matrix $C$ by column $l$, that is $O_{l}\,^{i}M_{p}=M_{p}^{i\rightarrow l}$, if
$p\leq i\leq p{+}k{-}1$ and $l<p$, while its action vanishes otherwise. As we
demonstrate in \appref{sec:Grass}, it is straightforward to show that
\begin{equation}
\sum_{l<i}
\Lambda^{\cA}_i\Lambda^{\cB}_l O_l\,^i M_{p}
=
\sum_{l=1}^{p-1}
\Lambda^{\cB}_l\Lambda^{\cA}_lM_{p}\,.
\end{equation}
In other words, the minors transform covariantly under the operator $O_l\,^i$.
Note that one must be careful as the integral formula has a $\grp{GL}(k)$
symmetry and is well defined only after gauge fixing. Thus, to prove the
invariance of the integral, one should either introduce a gauge-fixing
function, on which $O_{i}\,^{l}$ acts, or directly work with the gauge-fixed
integral.

For consistency, we will proceed with the gauge-fixed integral with the columns
$1$ through $k$ of the matrix $C$ set to the unit matrix:
\begin{equation}
\begin{pmatrix}
1 & \cdots & 0 & C_{1,k+1} & \cdots & C_{1,2k} \\
0 & \cdots & 0 & \vdots & \vdots & \vdots \\
0 & \cdots & 1 & C_{k+1,2k} & \cdots & C_{k,2k}
\end{pmatrix}
\,.
\end{equation}
We see that for $k\leq i$ the operator $O_{l}\,^{i}$ is defined simply by replacing
$C_{ai}\rightarrow C_{al}$
or $C_{ai}\rightarrow\delta_{al}$.
However, for $i<k$ the operator requires
careful treatment. For the four-dimensional case, this situation was
discussed in detail in~\cite{Drummond:2010qh}, where it was shown that
for $i\leq k$, $O_{l}\,^{i}$ should be replaced by
$\cN_{i}\,^{l}\equiv
\sum_{r=k+1}^{2k}C_{lr}\frac{\partial}{\partial C_{ir}}\,$, which is
nothing but a $\grp{GL}(k)$ rotation on the rows of the unfixed part
$C_{ai}$ of the
gauge-fixed $C$-matrix. Thus $\sum_{i<l}\cN_{i}\,^{l}M_{p}=0$ for
$k<p\leq 2k$, whilst $\cN_{i}\,^{l}M_{p}=-M_{p}^{l\rightarrow i}\,$ for
$p\leq k$.

Collecting these results, and noting that since the undeformed Gra{\ss}mannian
integral vanishes under $\gen{\widehat J}^{(\cA\cB]}$, we can
focus solely on the extra terms that are generated due to the additional
exponents $b_i$ of the measure. These additional terms are given by
\begin{equation}
\gen{\widehat
J}^{(\cA\cB]}\grass_{2k}(b_i)=
\grass_{2k}(b_i)
\sum_{j=1}^{2k}
\biggbrk{\sum_{l=j+1}^k 2b_l-b'_{j}}
\Lambda_{j}^{(\cA}\Lambda^{\cB]}_{j}\,,
\end{equation}
where $b'_j$ is defined as:
\begin{equation}
\lrbrk{\sum_{a=1}^k\,C_{aj}\frac{\partial}{\partial C_{aj}}}
\frac{1}{\prod_{i=1}^{k} M_i(C)^{b_i}}
=
b'_j\frac{1}{\prod_{i=1}^{k} M_i(C)^{b_i}}\,.
\end{equation}
Note that unlike in $\superN=4$ SYM theory, the eigenvalue $b'_j$ does not
correspond to a central charge.
We will further comment on this point below.
In terms of the exponents $b_i$, the eigenvalue $b'_j$ expands to
\begin{equation}
-b'_j=
\begin{cases}
b_1+\dots+b_j & j\leq k\\
b_{j-k+1}+\dots+b_k & j\geq k\,.
\end{cases}
\label{eq:bprime}
\end{equation}
To retain Yangian invariance, it is necessary to deform the
level-one generators by
\begin{equation}
\gen{\widehat J}^{(\cA\cB]}
\rightarrow
\gen{\widehat J}^{(\cA\cB]}
+\sum_{j=1}^{2k}u_j\,\Lambda_{j}^{(\cA}\Lambda^{\cB]}_{j}\,,
\end{equation}
where, for general $n=2k$, the relation between the deformation
parameters is given by
\begin{equation}
\label{ABJMmaster}
\sum_{l=j+1}^k2 b_l+u^{-}_j
=
\textrm{constant}\,.
\end{equation}
Here, we define $u_j^-=u_j-b'_j$, and the constant must be
independent of $j$.
This implies
\begin{equation}
\label{ABJMmaster2}
\half(u_j^--u_{j-1}^-)
=
\begin{cases}
b_j & \text{for } j\leq k\\
0   & \text{for } j>k\,.
\end{cases}
\end{equation}
Using \eqref{eq:bprime}, these conditions can be rewritten as
\begin{equation}
\label{ABJMmaster3}
\eqbox{
u_j=u_{j+k}\,,
\qquad
b_j=u_j-u_{j-1}\,,
\qquad
1\leq j\leq k\,.
}
\end{equation}
In particular, this reproduces \eqref{eq:u_perm} for the permutation
$\sigma$ of the top cell, which is just a cyclic shift by $k$ sites.
Note that \eqref{ABJMmaster,ABJMmaster2} closely resemble the
constraints \eqref{eq:cons4d,eq:topcellexp} of the four-dimensional
case. For four points, we have $b_1=\half(u_1^--u_{4}^-)$ and
$b_2=\half(u_2^--u_{1}^-)$, which, combined with $b_1+b_2=0$, implies
$u_4=u_2$ and $b_1=u_1-u_2$, in agreement with \eqref{eq:4ptsol}.

\paragraph{Little Group and Fermion Number.}

In our discussion of the invariance of the deformed Gra{\ss}mannian
integral, we encountered the scaling operator
$\mathfrak{f}_j=\Lambda_j^\cC\,\partial/\partial\Lambda_j^\cC$,
which acts on the external scattering data as
\begin{equation}
\label{eq:scalingop}
\mathfrak{f}_j
=
\lambda_j^{\alpha}\frac{\partial}{\partial\lambda_j^{\alpha}}
+\eta_j^A\frac{\partial}{\partial\eta^A_i}\,.
\end{equation}
This operator generates the three-dimensional little group $\mathbb{Z}_2$: The
exponentiated operator
\begin{equation}
\cF_j\equiv\exp{(i\pi\,\mathfrak{f}_j)}
\end{equation}
commutes with the whole $\alg{osp}(6|4)$ algebra, and the amplitude transforms according to
\begin{equation}
\cF_j\,\cA(\bar{1}2\bar{3}\dots 2k)
=(-1)^j\cA(\bar{1}2\bar{3}\dots 2k)\,.
\end{equation}
As pointed out in~\cite{Bargheer:2010hn}, this equation looks similar to the
local central charge constraint in $\superN=4$ SYM theory. The group-like
operator $\cF_j$ measures the fermion number, i.e.\ whether the
external leg $j$ is a bosonic or fermionic superfield $\Phi$ or $\bar\Phi$,
respectively.

An obvious question is how the deformed invariants behave under the operator
$\cF_j$. For the deformed four-point amplitude \eqref{eq:4ptMom} the
answer is simple and similar to the central charge constraint in four
dimensions: While the local invariance under $\cF_j$ is broken (only
for the fermionic legs though), the global constraint given by $\prod_j
\cF_j\amp_4=\amp_4$ is preserved. Here we consider the product of
$\cF_j$ due to the group-like structure of the operator $\cF$
as opposed to $\mathfrak{f}$.

Note that the superfield in $\superN=4$ SYM theory has a similar inconspicuous
symmetry under the operator
\begin{equation}
\op{F}^\text{4d}_j=\exp i\pi\biggbrk{\lambda_j^{\alpha}
\frac{\partial}{\partial
\lambda_j^{\alpha}}+\bar\lambda_j^{\dot\alpha}
\frac{\partial}{\partial\bar\lambda_j^{\dot\alpha}}+\eta_j^A
\frac{\partial}{\partial\eta_j^A}}\,:
\qquad
\op{F}^\text{4d}_j\amp_n=+\amp_n\,.
\end{equation}
This corresponds to the fact that $\Phi(-\Lambda)=\Phi(\Lambda)$,
i.e.\ to the statement that the total number of spinors in all terms
of the bosonic superfield is even, or even more simple: the bosonic
superfield is bosonic. Note that
$\op{F}^{\text{4d}}_j$ is not generated by the central charge
$\gen{C}_i$.
The breaking of the local fermion number operators $\cF_j$ and
$\cF^\text{4d}_j$ in three and four dimensions, respectively,
demonstrates the anyonic character of the above deformations.

\paragraph{The Cells of the Deformed Gra{\ss}mannian and BCFW.}

The integral in \eqref{Int} is a $k(k{-}1)/2$-dimensional integral
representing the top cell, and it is invariant under the deformed
Yangian. Here the evaluation parameters $u_i$ of the level-one
generators are constrained by \eqref{ABJMmaster3}. The bosonic delta function
$\delta^{2k}(C\cdot\lambda)$ imposes $(2k{-}3)$ constraints, and thus
the top cell has dimension $(k{-}2)(k{-}3)/2$. If some of the
deformation parameters are turned off, then one can localize the
top cell by residues on poles in the respective minor to obtain
lower-dimensional cells, and thus obtain deformed descendant
invariants. However, this does not yield the most general deformations
of the lower-dimensional cells. As shown in \secref{sec:glue}, one can
instead directly transform the lower-dimensional cells (on-shell
diagrams), which leads to further deformations that do not form
boundaries of the deformed top cell.

We can reconsider the question of consistent deformations
for the BCFW terms of tree amplitudes from the perspective of the
Gra{\ss}mannian integral. Similar to the four-dimensional case,
consistent deformations are only possible if the tree contour involves
residues on fewer than $(k-1)$ minors. As soon as the tree contour
includes poles from $(k-1)$ or all $k$ minors, all exponents
$b_i$ need to be set to zero and no deformation remains.
For six points, the amplitude
is the top cell, and thus there is a consistent two-parameter deformation.
For eight points, the BCFW terms are given by the sum of
residues for $M_1$ and $M_3$, and thus only a one-parameter
deformation remains. For ten points, as discussed in~\cite{Huang:2013owa},
the five BCFW terms are given by the zeros of
\begin{equation}
\{4,5,1\},\;\{5,1,2\},\;\{3,4,5\},\;\{2,3,4\},\;\{1,2,3\},
\end{equation}
where $\{i,j,k\}$ indicates the collection of minors that are
necessary to localize the three-dimensional integral. As one can see,
all five minors are involved in the localization, and no consistent
deformation remains. It is to be expected that there will be no
BCFW-preserving deformation for any amplitude beyond eight points.
These results are consistent with the
on-shell diagram analysis in \secref{sec:glue} above.

\subsection{R-matrix Construction for ABJM}
\label{sec:RmatABJM}

In \secref{sec:glue} above, we have verified the invariance of the
deformed four-point amplitude $\amp_4(z)$ under the evaluation
representation of the Yangian generators, and have obtained higher-point
diagrams by successive gluing of the fundamental four-point invariant.
In this section, we will identify $\amp_4(z)$ with
an integral kernel for the R-matrix $R_{jk}(z)$ of an integrable model, where
$z$ represents the spectral parameter.

\paragraph{R-Matrix.}

Let us define the action of the operator $R_{jk}(z)$ on
a function $f(\Lambda)$ by%
\footnote{Since $\OG_2$ has two branches, we correspondingly have
two different R-matrices $R^{\pm}(z)$.
In the following we concentrate on one of the branches, say $R^{+}(z)$.
Note that the actual undeformed 4-point amplitude is a linear combination of two contributions
from two kinematical branches $R^{\pm}(z)$,
each contribution being separately Yangian-invariant
(see, however, the comments on
the discussion of the collinear anomaly in Section~\ref{sec:discussion}).}
\begin{equation}
(R_{jk}(z)\circ f)(\dots,\Lambda_j,\Lambda_k,\dots)
\equiv\int\!\dd\Lambda_{\sharp}\,\dd\Lambda_{\flat}\,
\amp_4(z)(\Lambda_j,\Lambda_k, i\Lambda_{\sharp}, i\Lambda_{\flat})\,
f(\dots,\Lambda_{\flat},\Lambda_{\sharp},\dots)
\,.
\label{eq:Rabjm}
\end{equation}
An important property of this operator $R_{jk}(z)$ is that it
preserves the Yangian invariance when applied to a Yangian-invariant function.
To show this, first note that the expression \eqref{eq:Rabjm}
can be rewritten as
\begin{multline}
(R_{jk}(z)\circ f)(\dots,\Lambda_j,\Lambda_k,\dots)
=
\int\dL_{\sharp}\,\dL_{\flat}\,\dL_{\natural}\,\dL_{\diamond}
\\
\delta^{2|3}(\Lambda_{\natural}-i\Lambda_{\sharp})\,
\delta^{2|3}(\Lambda_{\diamond}-i\Lambda_{\flat})\,
\amp_4(z)\brk{\Lambda_j,\Lambda_k,\Lambda_{\natural},\Lambda_{\diamond}}\,
f(\dots,\Lambda_{\flat},\Lambda_{\sharp},\dots)\,.
\label{eq:rfexp}
\end{multline}
The expression \eqref{eq:rfexp} is a combination of the Yangian-preserving operations
discussed in Section~\ref{sec:glue}: We first take the product
of two Yangian invariants, $\cA_4(\Lambda_j,\Lambda_k,
\Lambda_{\natural},\Lambda_{\diamond})$ and
$f(\Lambda_{\flat},\Lambda_{\sharp})$, and then glue these objects by using the two
delta-function identifications $\Lambda_{\natural}=i\Lambda_{\sharp}$ and
$\Lambda_{\diamond}=i\Lambda_{\flat}$.
This implies that $R_{jk}(z)$ as defined in \eqref{eq:Rabjm} preserves the Yangian invariance.
Recall that
$\amp_4(z)\brk{\Lambda_j,\Lambda_k,\Lambda_{\natural},\Lambda_{\diamond}}$
is invariant under the Yangian with evaluation
parameters that satisfy $u_j=u_{\natural}$, $u_k=u_{\diamond}$, and
$z=\pm(u_j-u_k)$ according to \eqref{eq:fig4vertex}.
In addition, the gluing conditions \eqref{eq:gluecond} require
$u_\natural=u_\sharp$ and $u_\diamond=u_\flat$. Hence the action of
$R_{jk}$ permutes the evaluation parameters $u_j$ and $u_k$: If
$f(\dots,\Lambda_j,\Lambda_k,\dots)$ is Yangian invariant with
$\vec u=(\dots,u_j,u_k,\dots)$, then
$\brk{R_{jk}\circ f}(\dots,\Lambda_j,\Lambda_k,\dots)$ is invariant with
$\vec u=(\dots,u_k,u_j,\dots)$.
In other words, we have
\begin{equation}
[\gen{J}^a, R_{jk}(z)]=0\,,
\qquad
\widehat{\gen{J}}^a(\dots,u_j,u_k,\dots)\,R_{jk}(z)=R_{jk}(z)\,\widehat{\gen{J}}^a(\dots,u_k,u_j,\dots),
\label{eq:R_invariance}
\end{equation}
when acting on Yangian-invariant functions. Here the number of legs in the
definition of the generators $\gen{J}^a$ and $\gen{\widehat J}^a$
\sbrk{cf.\ \eqref{eq:lev1_J}} depends on the number of legs of the invariant acted on.
However, since all other terms commute trivially, the Yangian
generators $\gen{J}^a$ and $\widehat{\gen{J}}^a$ in
\eqref{eq:R_invariance} reduce to the two-site
Yangian generators with evaluation parameters
$(u_j,u_k)$.
Note that invariance is only preserved when $R_{jk}$ acts on
adjacent legs of the invariant $f$.

For later purposes, let us simplify the definition of the R-matrix.
Plugging in the definition of the four-point amplitude in \eqref{eq:4vertint}, we obtain
\begin{equation}
(R_{jk}(z)\circ f)(\Lambda_j,\Lambda_k)
=
\int\frac{\dd\theta}{\sin(\theta)^{1+z}}
\int\dL_{\sharp}\,\dL_{\flat}\,
\delta^{4|6}\bigbrk{C(\theta)\cdot(\Lambda_j,\Lambda_k, i\Lambda_{\sharp}, i\Lambda_{\flat})}\,
f(\Lambda_{\flat},\Lambda_{\sharp})\,,
\label{eq:Rf_comp}
\end{equation}
where the matrix $C(\theta)$ is defined as in \eqref{eq:Cmatrix_4pt}.
We can trivially solve the delta-function constraint for $\Lambda_{\sharp},
\Lambda_{\flat}$, giving rise to
\begin{equation}
\label{Rmatrix1}
(R_{jk}(z)\circ f)(\Lambda)
\equiv
\int\frac{\dd\theta}{\sin(\theta)^{1{+}z}}
f(\Lambda)\raisebox{-.2em}{\Big|}_{\begin{subarray}{l}
\Lambda_j\to\,+\sin(\theta)\Lambda_k+\cos(\theta)\Lambda_j,\\
\Lambda_k\to\,-\cos(\theta)\Lambda_k+\sin(\theta)\Lambda_j
\end{subarray}}\,.
\end{equation}
%

\paragraph{RLL Relation.}

\begin{figure}
\centering
\includegraphicsbox{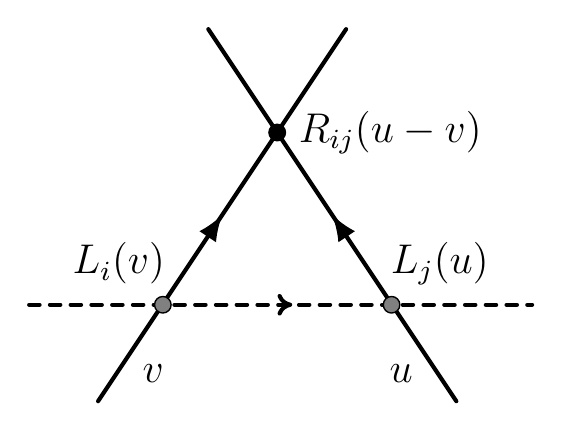}
\quad{\Large $=$}\quad
\includegraphicsbox{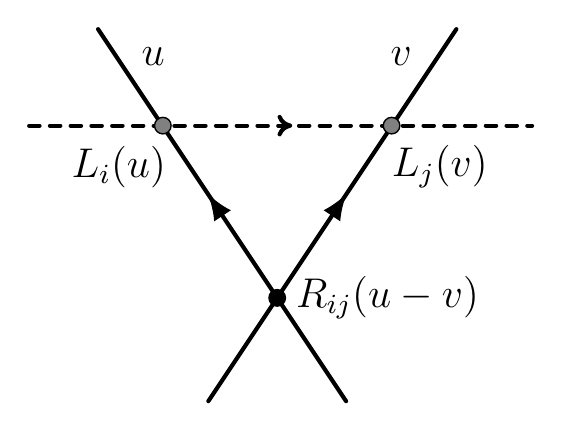}
\caption{The graphical representation of the RLL relation.
The R-matrix is associated with an intersection of two undotted lines, while the L-matrix with
that of an undotted line and a dotted line.
The spectral parameters are associated with the particle lines, which will be
identified with the rapidity lines of
integrable models.}
\label{fig:RLL}
\end{figure}

The discussion in the previous sections can be nicely reformulated in the language of
integrable models.
To explain this, let us first define the L-operator $L_i(u)$ by
\begin{equation}
L_i(u)\equiv u\textbf{1}+\sum_a\gen{J}_i^a\,\bfm{e}_a\,,
\label{eq:Lax}
\end{equation}
where $\gen{J}_i^a$ are the level-zero generators for the representation
of the particle $i$, and
$\bfm{e}_a$ denotes the generators of the fundamental representation.
Let us also define the monodromy operator by
\begin{equation}
T(u_0,\vec{u})
\equiv
L_{1}(u_0-\half u_{1})
L_{2}(u_0-\half u_{2})
\dots
L_{2k}(u_0-\half u_{2k}).
\label{eq:Tdef}
\end{equation}
By standard procedure,%
\footnote{See e.g.~\cite{Dolan:2004ys}.}
expanding the monodromy
yields the Yangian generators:
\begin{equation}
T(u_0,\vec{u})=\sum_{n=0}^{2k} u_0^{2k-n}\,\gen{J}^{(n-1)} (\vec{u})\,,
\label{eq:T_expand}
\end{equation}
where $\gen{J}^{(n)}(\vec{u})$ is (up to additive constants and
combinations of
lower-level generators) the level-$n$ generator
with evaluation parameters $\vec{u}$, namely%
\footnote{Recall the
constraint $\sum_iu_i=0$.}
\begin{equation}
\gen{J}^{(-1)}=\bfm{1}\,,
\quad
\gen{J}^{(0)}(\vec{u})=\gen{J}^a\bfm{e}_a\,,
\quad
\gen{J}^{(1)}(\vec{u})
=\half\bigbrk{
\widehat{\gen{J}}^a\bfm{e}_a
+\gen{J}^a\bfm{e}_a\,\gen{J}^b\bfm{e}_b
-\alpha\,\gen{J}^a\bfm{e}_a
+\half\sum_{i<j}u_iu_j\bfm{1}}\,.
\label{eq:jfromt}
\end{equation}
Here, the constant $\alpha$ stems from the single-site relation
\begin{equation}
\gen{J}_i^a\bfm{e}_a\,\gen{J}_i^b\bfm{e}_b=\alpha\,\gen{J}_i^a\bfm{e}_a\,.
\label{eq:alpharel}
\end{equation}
This relation is representation-dependent, but holds for the
fundamental representation. It ensures that the Yangian generators
obey the Serre relations~\cite{Dolan:2004ps,Bargheer:2010hn}.
Now we can use \eqref{eq:T_expand} to encode \eqref{eq:R_invariance}
into the ``RLL'' relation
\begin{equation}
R_{ij}(u_j-u_i) L_i(u_0-\half u_i) L_j(u_0-\half u_j)
=
L_i(u_0-\half u_j) L_j(u_0-\half u_i) R_{ij}(u_j-u_i)\,,
\label{eq:RLL}
\end{equation}
see \figref{fig:RLL}.
This equation is the so-called RLL relation,
which is one version of the Yang--Baxter relation often found in
integrable
models~\cite{Takhtajan:1979iv,Kulish:1981gi,Kulish:1981bi,Faddeev:1996iy}.
As we have seen, this relation encodes the fact that $R_{ij}(z)$ preserves the Yangian invariance.
The relation \eqref{eq:RLL} holds
when the operators act in the space of Yangian-invariant functions;
this will be sufficient for the construction of Yangian invariants in
the later part of this section.%
\footnote{We thank Carlo Meneghelli for helpful comments on the
relation between \eqref{eq:RLL} and \eqref{eq:R_invariance}.}

\paragraph{Yang--Baxter Equation.}

The relation \eqref{eq:RLL} means that the R-operator is the
intertwiner for the tensor product of representations of
$Y[\alg{osp}(6|4)]$.
In particular, consistency with the associativity of the tensor product is guaranteed
by the Yang--Baxter relation:
\begin{equation}
R_{ij}(w-v)R_{j\ell}(w-u)R_{ij}(v-u)
=
R_{j\ell}(v-u)R_{ij}(w-u)R_{j\ell}(w-v)\,.
\label{Yang-Baxter}
\end{equation}
For our R-matrix \eqref{Rmatrix1},
this can be shown to hold by direct computation: The $\theta$-rotation in
\eqref{Rmatrix1} is a rotation in the
$(\Lambda_i,\Lambda_j)$-plane, and both sides of \eqref{Yang-Baxter} give rise
to a parametrization of the
rotation group in terms of Euler angles, which are thus related by a coordinate transformation.
One can verify explicitly that the product of measure factors
in the integrals is kept invariant by the transformation.

This result shows that the R-operator \eqref{Rmatrix1}
gives the R-matrix for a representation of $Y[\alg{osp}(6|4)]$.
It can be written in the Gra{\ss}mannian integral form, a fact which we have
not found in the literature.
It would be nice to compare our expression for the R-matrix
with the known expressions in the literature.

\paragraph{Yangian Invariants.}

Having understood the four-vertex, the next task is to understand the more
complicated on-shell diagrams obtained by fusing lines.
In the language of R-matrices, this can be reformulated as the statement that
higher-point invariants are
obtained by iterated action with the R-matrix on vacuum delta functions.
This is similar to the $\superN=4$ case \sbrk{see the discussion
around
\eqref{Arecursive}}.%
\footnote{The connection to Yangians and integrable models is more direct for
ABJM theory than for the $\superN=4$ theory.
For ABJM theory, the R-matrix \eqref{eq:Rabjm} coming from the BCFW shift
directly gives the R-matrix for (a representation of) the Yangian.
By contrast the operator coming from the BCFW shift in $\superN=4$ SYM is the 3-point amplitude,
while the R-matrix for the Yangian corresponds to a 4-point amplitude,
which is obtained by combining four BCFW operators.}

To explain this, let us start with an on-shell diagram,
i.e.\ a set of $k$ lines connecting $2k$ points on a circle,
such that
no three lines intersect at the same point.
By following each line, we obtain a permutation $\sigma$ of order two,
$\sigma^2=1$. That is, $\sigma$ decomposes into $k$
commuting transpositions of two elements:
$\sigma=\sigma_k\dots\sigma_1$, with each $\sigma_j=[a_{j}, b_{j}]$.
Just as the diagram itself, also
the permutation is kept invariant under triangle moves.
While every diagram has a unique associated permutation, the converse
is only true for reduced diagrams. Every permutation of order two
uniquely specifies a reduced diagram (modulo triangle moves), but the
same permutation is associated to an infinite number of inequivalent unreduced diagrams.%
\footnote{Among all the diagrams associated to a given permutation,
the reduced diagram is the one that has minimal degree (number of
integration variables).}

Now, every on-shell diagram can be obtained from a $k$-line diagram without any
four-vertex (a ``vacuum diagram'')
by a sequence of BCFW bridges~\cite{ArkaniHamed:2012nw,Huang:2013owa}.
Let us first restrict to reduced diagrams, which are uniquely
specified by their permutation.
Starting from the ``vacuum permutation''%
\footnote{The choice of vacuum is not unique, but one can restrict to
this particular choice~\cite{Huang:2013owa}.}
\begin{align}
\label{eq:sigma_vac}
\sigma_{\rm vac}=[12][34]\dots[2k-1,2k]\,,
\end{align}
we can arrive at any other permutation $\sigma$ (representing a reduced on-shell diagram)
by a sequence of BCFW bridges:
Each BCFW bridge lets two adjacent legs intersect, and thus conjugates
the associated permutation by a transposition.
Hence
\begin{equation}
\sigma=\sigma\indup{R}\,\sigma\indup{vac}\,\sigma\indup{R}^{-1}\,,
\qquad
\sigma\indup{R}=[i_{\ell},j_{\ell}]\dots[i_{1},j_{1}]\,,
\label{eq:sigma_decomp}
\end{equation}
with each $[i_m,j_m]$ being a transposition
of two \emph{adjacent} elements.

Let us translate this into the language of integrable models.
With the vacuum permutation $\sigma\indup{vac}$
we associate an amplitude
which is given by a product of delta functions
\begin{equation}
\Omega_{2k}
\equiv\prod_{j=1}^k\delta^{2|3}(\Lambda_{2j-1}+ i\Lambda_{2j})
\,.
\label{eq:vacabjm}
\end{equation}
Next, each BCFW bridge
is represented by the R-matrix acting on the two respective lines,
as we have already seen. This means that
a general Yangian invariant, corresponding to a general reduced on-shell
diagram described by $\sigma$, can be obtained by acting with a chain
of R-matrices on the vacuum amplitude:
\begin{equation}
\cY_{\sigma}(z_1,\dots,z_{\ell})
=
R_{\sigma\indup{R}}(\vec{z})\,\Omega_{2k}
=
R_{i_\ell, j_\ell}(z_\ell)\dots R_{i_1, j_1}(z_1)\,\Omega_{2k}\,,
\label{eq:YangianABJM}
\end{equation}
where the sequence of transpositions $\sigma\indup{R}=[i_{\ell},
j_{\ell}]\dots[i_{1}, j_{1}]$ is defined by \eqref{eq:sigma_decomp}.

For a given permutation $\sigma$ (of order two), both the choice of
$\sigma\indup{R}$ and its decomposition into adjacent transpositions
are not unique. Two permutations $\sigma\indup{R}$, $\sigma\indup{R}'$
lead to the same permutation $\sigma$ if and only if
$\sigma\indup{R}'=\sigma\indup{R}\,\sigma'$, where $\sigma'$ is in the
centralizer $\grp{C}(\sigma\indup{vac})$. Hence the distinct
permutations $\sigma$ are in one-to-one correspondence with the
elements of the coset $\grp{S}_n/\grp{C}(\sigma\indup{vac})$.
The ambiguity in the decomposition of $\sigma\indup{R}$ into adjacent
transpositions is due to the permutation
group relation
\begin{equation}
[i,i+1][i+1,i+2][i,i+1]=[i+1,i+2][i,i+1][i+1,i+2]\,.
\end{equation}
In terms of invariants \eqref{eq:YangianABJM}, this identity amounts
to the triangle move alias Yang--Baxter equation \eqref{Yang-Baxter}.
Also the ambiguity in the choice of $\sigma\indup{R}$ is due to this
relation, in this case applied to the full permutation $\sigma$.
Hence, for a given permutation $\sigma$, the invariant
\eqref{eq:YangianABJM} is independent of the choice and decomposition
of $\sigma\indup{R}$.

To summarize: For every reduced on-shell diagram,
there is a decomposition of the associated permutation into
adjacent transpositions that encodes the chain of R-matrices that need
to act on the appropriate vacuum to reconstruct the
diagram. Even though the decomposition into transpositions is
ambiguous, the invariant is unique.

\emph{Unreduced on-shell diagrams} are not uniquely specified by their
associated permutation. Nevertheless, they are constructed just as
reduced diagrams, by successively applying BCFW bridges. Hence also
unreduced diagrams can be written as a chain of R-matrices that act on
a vacuum amplitude,
\begin{equation}
\cY_{[i_\ell,j_\ell],\dots,[i_1,j_1]}(z_1,\dots,z_{\ell})
=
R_{i_{\ell}, j_{\ell}}(z_{\ell})\dots R_{i_1, j_1}(z_1)\,\Omega_{2k}\,.
\label{eq:unredRchain}
\end{equation}
Here, the sequence of transpositions
$[i_{\ell},j_{\ell}],\dots,[i_{1},j_{1}]$ is sufficient to define the
diagram, even though the resulting permutation
$\sigma=[i_{\ell},j_{\ell}]\dots[i_{1},j_{1}]$ is not.

Our claim here is that \eqref{eq:YangianABJM,eq:unredRchain}
are indeed Yangian invariant if the
spectral parameters $z_i$ are constrained to obey the relation
\eqref{eq:fig4vertex}, that is
\begin{equation}
z_m=\pm(u_{j_m}-u_{i_m})\,.
\label{eq:Rzcond}
\end{equation}
Since the overall shift of $u_i$ is irrelevant,
only $(k-1)$ of these parameters are independent.

\paragraph{Yangian Invariance.}

Let us prove the Yangian invariance of \eqref{eq:YangianABJM}.
Note that the invariance already
follows from the fact that the action of an R-matrix is equivalent to
gluing an invariant four-vertex to another invariant (\secref{sec:glue}).
The purpose of this section is to recast the argument in a form closer to
standard integrable models.

\begin{figure}
\centering
\includegraphicsbox{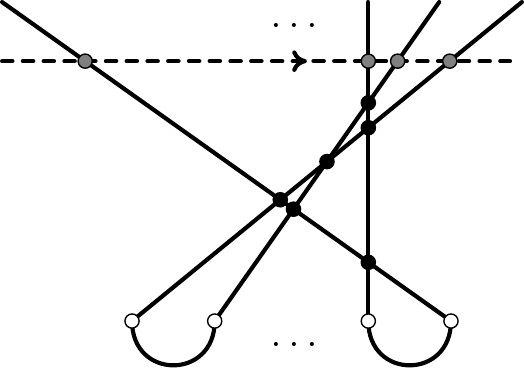}
\quad{\Large $=$}\quad
\includegraphicsbox{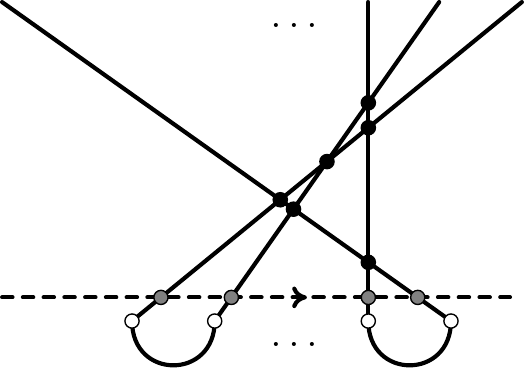}
\caption{Graphical proof of the Yangian invariance, c.f.\ \protect\eqref{eq:YangianProof}.
Using the RLL relations of \protect\figref{fig:RLL} multiple times, we
can commute the product of L-operators (gray dots) through the product
of R-matrices (black dots) when acting on the vacuum
\protect\eqref{eq:vacabjm} (half circles).
In the above example we have $k=2$, i.e.\ $2k=4$ external points, and the Yangian invariant
\protect\eqref{eq:YangianABJM} contains six R-matrices. The ``$\dots$''
represent additional lines that can be added in general.}
\label{fig:MLL}
\end{figure}

Since the monodromy operator is the generating function of the Yangian generators,
Yangian invariance is equivalent to the statement that
\eqref{eq:YangianABJM} is an eigenfunction
of the monodromy operator:
\begin{equation}
T(u_0,\vec{u})\,\cY_{\sigma}(\vec{z})
=
\prod_{i=1}^{2k}(u_0-\half u_i)\,\cY_{\sigma}(\vec{z})\,.
\label{eq:new_2}
\end{equation}
where $u_0$ is arbitrary and $\vec{u}$ are fixed to be the evaluation
parameters of the Yangian representation.
On the left-hand side we can express $T(u_0,\vec{u})$ as a product of the L-operators,
and then, due to \eqref{eq:Rzcond},
commute with the R-matrices with the help of the RLL relation \eqref{eq:RLL}:
\begin{align}
L_1(u_0-\half u_{1})\dots L_{2k}(u_0-\half u_{2k})\,R_{ab}(\vec{z})\,
=R_{ab}(\vec{z})\,L_{1}(u_0-\half u_{[ab](1)})\dots L_{2k}(u_0-\half u_{[ab](2k)})\,
\end{align}
for any transposition $[ab]$ of two adjacent elements.%
\footnote{Note that this argument does not work for the transposition
$[2k,1]$, since the RLL relation does not apply in this case. However,
one can always choose $\sigma\indup{R}$ such that it does not act on
the first and last legs (for any permutation $\sigma$). Hence at first
sight, \eqref{eq:YangianProof} only applies for such
$\sigma\indup{R}$. However, different choices of $\sigma\indup{R}$ are
related to each other by triangle moves, and we know that triangle
moves preserve diagrams, and hence also preserve Yangian invariance.
Therefore, $R_{\sigma\indup{R}}\Omega_{2k}$ is invariant for all
$\sigma\indup{R}$. In fact, from the on-shell diagram point of view,
the choice of first and last leg in the
definition \eqref{eq:Tdef} of the monodromy matrix is arbitrary, and
thus the invariance discussion should not depend upon this choice. Indeed
one can show that the Yangian algebra is invariant under cyclic
rotations of the chain of L-operators in \eqref{eq:Tdef}
(for algebras with vanishing dual Coxeter number).}
By induction we obtain (see
\figref{fig:MLL} for a graphical representation)
\begin{align}
\label{eq:YangianProof}
\begin{split}
T(u_0,\vec{u})\,\cY_{\sigma}(\vec{z})
&=T(u_0,\vec{u})\,R_{\sigma\indup{R}}(\vec{z})\,\Omega_{2k}\\
&=L_1(u_0-\half u_{1})\dots L_{2k}(u_0-\half u_{2k})\,R_{\sigma\indup{R}}(\vec{z})\,\Omega_{2k}\\
&=R_{\sigma\indup{R}}(\vec{z})\,
L_1(u_0-\half u_{\sigma\indup{R}(1)})\dots L_{2k}(u_0-\half u_{\sigma\indup{R}(2k)})\,\Omega_{2k}\,.
\end{split}
\end{align}
Recalling the definition \eqref{eq:vacabjm} of $\Omega_{2k}$, this
becomes
\begin{equation}
T(u_0,\vec{u})\,\cY_{\sigma}(\vec{z})
=
R_{\sigma\indup{R}}(\vec{z})
\prod_{j=1}^{k}
L_{2j-1}(u_0-\half u_{\sigma\indup{R}(2j-1)})\,
L_{2j}(u_0-\half u_{\sigma\indup{R}(2j)})\,
\delta^{2|3}\brk{\Lambda_{2j-1}+i\Lambda_{2j}}\,.
\label{eq:TonR}
\end{equation}
After some algebra, and using the relation \eqref{eq:alpharel}, each
factor in the product takes the form
\begin{multline}
L_i(u)L_j(v)\,\delta^{2|3}\brk{\Lambda_{i}+i\Lambda_{j}}
\\=
\Bigsbrk{
uv
+\bigbrk{u\,\gen{J}_j^a+v\,\gen{J}_i^a}\bfm{e}_a
+\half\gen{J}^a\bfm{e}_a\,\gen{J}^b\bfm{e}_b
-\half\alpha\,\gen{J}^a\bfm{e}_a
+\half\gen{J}_i^a\,\gen{J}_j^b\,f_{ab}{}^c\,\bfm{e}_c
}
\delta^{2|3}\brk{\Lambda_{i}+i\Lambda_{j}}\,,
\end{multline}
where $\gen{J}^a=\gen{J}_i^a+\gen{J}_j^a$ are the two-site level-zero
generators. Now the third and fourth term in the bracket vanish by
\eqref{eq:pi}, and the last term vanishes due to the argument below
\eqref{eq:fusion1}: It is proportional to the dual Coxeter number,
which is zero for $\alg{osp}(6|4)$. Finally, the vanishing of the
second term requires $u=v$. Hence,
\eqref{eq:TonR} equals \eqref{eq:new_2} if and only if
\begin{equation}
\eqbox{
u_{\sigma\indup{R}(2j-1)}=u_{\sigma\indup{R}(2j)}\,,
\qquad
j=1,\dots,k\,.
}
\label{eq:Rinvcond}
\end{equation}
Using \eqref{eq:sigma_decomp,eq:sigma_vac}, one can easily see that
these conditions are equivalent to the previously derived invariance
constraints \eqref{eq:u_perm}.
In conclusion, Yangian invariance of \eqref{eq:YangianABJM} is
recovered provided that the constraints \eqref{eq:Rinvcond} and
\eqref{eq:Rzcond} hold.

\paragraph{An Example.}

Let us consider as an example the deformation of the six-point top cell.
Like all other on-shell diagrams, it can be represented as a disk with
intersecting lines that end at the boundary, e.g.
\begin{equation}
\includegraphicsbox{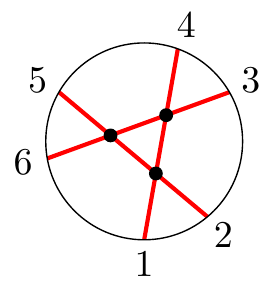}
\quad\sim\quad
\includegraphicsbox{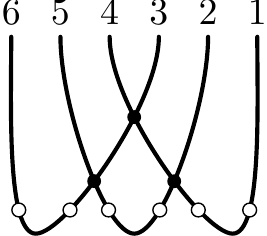}
\quad,
\end{equation}
where the second figure illustrates the relation to diagrams of the
type shown in \figref{fig:MLL}.
This diagram corresponds to a permutation $\sigma=[14][25][36]$.
As described above, we can build such a diagram from a vacuum diagram
by applying successive transpositions:
\begin{equation}
\label{eq:Sequence1}
\includegraphicsbox{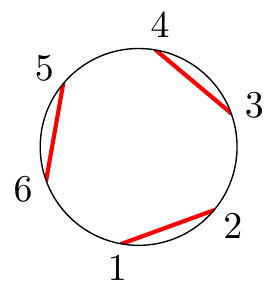}
\xrightarrow{\ [45]\ }
\includegraphicsbox{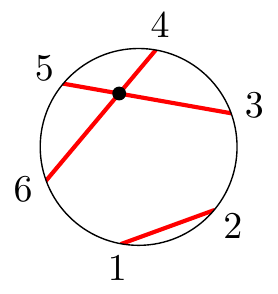}
\xrightarrow{\ [23]\ }
\includegraphicsbox{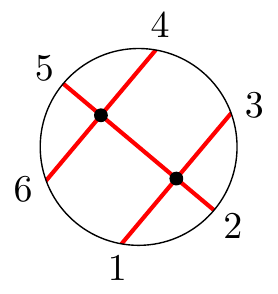}
\xrightarrow{\ [34]\ }
\includegraphicsbox{Fig6pt4.pdf}\,.
\end{equation}
Above, each transposition labeled by
$[ij]$ corresponds to applying
an additional four-point vertex\,/\,R-matrix to the previous diagram.
For the permutation, this means
\begin{equation}
\sigma=[34][23][45]\sigma_{\rm vac}[45][23][34]\,,
\quad
\sigma_{\rm vac}=[12][34][56]\,.
\end{equation}
The vacuum amplitude \eqref{eq:vacabjm} is simply given by
\begin{equation}
\Omega_6
=
\delta^{2|3}(\Lambda_1+i\Lambda_2)\,
\delta^{2|3}(\Lambda_3+i\Lambda_4)\,
\delta^{2|3}(\Lambda_5+i\Lambda_6)\,.
\end{equation}
We can then act iteratively with the R-operator \eqref{Rmatrix1} and
the sequence in \eqref{eq:Sequence1} translates into
\begin{equation}
\label{eq:RRR2}
R_{34}(z_3)\,R_{23}(z_2)\,R_{45}(z_1)\,\Omega_6\,.
\end{equation}
The corresponding orthogonal Gra{\ss}mannian $C$-matrix takes the form:
\begin{equation}
\begin{pmatrix}
1 & i\mathrm{c}_2 & i\mathrm{s}_2\mathrm{c}_3                            & i\mathrm{s}_2\mathrm{s}_3                           & 0             & 0 \\
0 & \mathrm{s}_2  & -\mathrm{c}_2\mathrm{c}_3 +i\mathrm{c}_1\mathrm{s}_3 & -i\mathrm{c}_1\mathrm{c}_3-\mathrm{c}_2\mathrm{s}_3 & i\mathrm{s}_1 & 0 \\
0 & 0             & \mathrm{s}_1\mathrm{s}_3                             & -\mathrm{s}_1\mathrm{c}_3                           & -\mathrm{c}_1 & i
\end{pmatrix}\,,
\end{equation}
where $\mathrm{s}_j=\sin\theta_j$ and $\mathrm{c}_j=\cos\theta_j$.
One can verify that this indeed corresponds to the top-cell diagram,
as all consecutive minors are non-vanishing.

As we have seen above, diagrams that are equivalent under triangle moves simply
correspond to using a different sequence of R-matrices that yields the
same final permutation. In our example, the triangle move corresponds
to the equivalence between \eqref{eq:RRR2} and
\begin{equation}
\label{eq:RRR1}
R_{61}(z_3)R_{23}(z_2)R_{45}(z_1)\Omega_6\,,
\end{equation}
The
corresponding matrix parametrizing the orthogonal Gra{\ss}mannian is given by
\begin{equation}
\begin{pmatrix}
-c_3           & i\mathrm{c}_2 & i\mathrm{s}_2 & 0             & 0             & \mathrm{s}_3 \\
0              & \mathrm{s}_2  & -\mathrm{c}_2 & i\mathrm{c}_1 & i\mathrm{s}_1 & 0 \\
-i\mathrm{s}_3 & 0             & 0             & \mathrm{s}_1  & -\mathrm{c}_1 & i\mathrm{c}_3
\end{pmatrix}\,.
\end{equation}
Again one can verify that all adjacent minors are non-vanishing, and hence
the top cell is recovered.

\section{Discussion, Conclusions \& Outlook}
\label{sec:discussion}

In this paper, we have considered integrable deformations of
scattering amplitudes in four-dimensional $\superN=4$ super
Yang--Mills theory and three-dimensional ABJM theory. We found a similar
structure of deformed invariants for both theories, which incorporates the
deformed Gra{\ss}mannian integrals, the construction of deformed on-shell
diagrams via gluing, as well as the algebraic R-matrix construction.
\medskip

Interestingly, while part of the deformation parameters in four
dimensions is associated with the violation of invariance
under the local central charge generators of $\superN=4$ SYM theory, there is
no such central charge for the three-dimensional symmetry algebra
$\alg{osp}(6|4)$. Furthermore, the four-dimensional deformations remind
of deformed helicities, but the external particles in ABJM do not carry
helicity charges either. Nevertheless, we demonstrate that consistent
deformations are possible in 3d ABJM theory, which can be attributed to
the introduction of non-trivial evaluation parameters in the Yangian
level-one generators. A local operator, similar to the central charge
generator in four dimensions, is the
$\mathbb{Z}_2$ phase of the three-dimensional little group in ABJM.
We have briefly commented on this fermion number operator at the end of
\secref{sec:glue}, whose breaking indicates the anyonic nature of the
introduced deformations.
We close with
some comments on both cases, and discuss future directions.
\medskip

\begin{table}[t]
\centering
\begin{tabular}{|c|c|c|}\hline
Yangian invariant&$\superN=4$ SYM Theory& $\superN=6$ SCS Theory\\\hline\hline\rule{0pt}{12pt}
$n$-leg amplitude&MHV, \dots, $\overline{\text{MHV}}$& \\\hline
3&2, 2&---\\
4&3&1\\
5&4, 4&---\\
6&5, 1, 5&2\\
7&6, 0, 0, 6&---\\
8&7, 0, 0, 0, 7&1\\
9&8, 0, 0, 0, 0, 8&---\\
$n\geq10$&$n-1$, 0, \dots, 0, $n-1$& 0
\\\hline\hline
$n$-leg diagram&$n-1$&$\frac{n}{2}-1$\rule{0pt}{10pt}
\\\hline
\end{tabular}
\caption{Comparison of the deformation degrees of freedom. Each
Yangian-invariant $n$-leg diagram in four or three dimensions has $n-1$ or
$\sfrac{n}{2}-1$ free parameters, respectively. Scattering amplitudes on the
other hand may be BCFW-composed of several diagrams. Requiring that the
external data of all diagrams contributing to a certain amplitude is the same,
this imposes stronger constraints, which in general result in less degrees of
freedom. In four dimensions the numbers were checked explicitly up to $n=16$.
The numbers in ABJM, and at higher $n$ in four dimensions, result from the
naive counting of degrees of freedom and constraints; for larger $n$ the
constraints outweigh the parameters and no degrees of freedom remain (beyond
the MHV sector in 4d).
}
\label{tab:comptable}
\end{table}

Certainly the most pressing question is whether the deformations
discussed here will be useful for computing loop amplitudes. Up to
now, the deformations might mostly look like a mathematical
curiosity. For instance, the famous BCFW decomposition cannot be deformed
consistently, not even at tree level~\cite{Beisert:2014qba}, cf.\ \tabref{tab:comptable}.%
\footnote{In principle, one could consider deforming each term in the
BCFW decomposition with a different set of central charges $c_i$, as
long as the Yangian evaluation parameters $u_i$ remain universal.
Empiric case studies for higher multiplicities (up to $n=18$) and
helicities (up to $k=5$) suggest that such deformations are admissible
for generic amplitudes. However, while a plethora of deformation
parameters remains unconstrained, the physical interpretation and the
practical use of such deformations remains unclear.}
Moreover, the deformed one-loop four-point amplitude in four dimensions generically
integrates to zero: Only very special deformations give a
non-vanishing result~\cite{Beisert:2014qba}. Still there are a few
interesting approaches one might want to pursue.
For example, noting that the four-point amplitude is
maximally helicity violating, it is not excluded that suitable deformations will
be useful for computing the ratio function of~\cite{Drummond:2008vq}.
Perhaps most promising is the idea to give up on a BCFW-like
decomposition, and to interpret the deformed top cell, or equivalently the
deformed Gra{\ss}mannian integral, as the complete deformed amplitude. The
challenge is to find a suitable
contour on which the deformed top cell integrates to a useful function.
As proposed in~\cite{Ferro:2014gca}, one could try to
require that the integrated result is meromorphic in the deformation
parameters. Interestingly, irrespective of the contour, the resulting
function will have deformed helicities for the external legs. This
points towards a possible connection with continuous-spin theories
proposed recently~\cite{Schuster:2013pxj,Schuster:2014xja}.
\medskip

Note that the form of the (deformed) Gra{\ss}mannian integrals mainly
follows from the symmetry structure of the underlying gauge theory.
This suggests to identify a similar Gra{\ss}mannian integral also in
other theories, for instance in two and six dimensions, where much
less is known about scattering amplitudes. In particular, for the
two-dimensional theories with an AdS${}_3$ string dual, it would be
interesting to initiate the study of scattering amplitudes based on
the symmetry algebras in analogy to the steps taken in~\cite{Bargheer:2010hn}
for three dimensions. Comparison should then
allow us to write down a Gra{\ss}mannian integral and to study
amplitude-like symmetry invariants. This could be helpful to make
progress on understanding the gauge-theory duals of these
AdS${}_3$ string theories.
\medskip

Importantly, here---as in all previous considerations of the deformed
scattering invariants---we have not considered the exact symmetry
generators of $Y[\alg{psu}(2,2|4)]$ and $Y[\alg{osp}(6|4)]$. That is
we have ignored the fact that at collinear momentum configurations the
above symmetry generators do not annihilate the tree-level S-matrices, but have
to be corrected due to the collinear
anomaly~\cite{Bargheer:2009qu,Beisert:2010gn,Bargheer:2011mm,Bargheer:2012cp,Bianchi:2012cq}.
It is known that these collinear contributions recursively relate
amplitudes with different numbers of external legs to each other and
it might be very enlightening to see whether these relations impose
further constraints on the deformation parameters. In this context, it
is interesting to note that the vacuum \eqref{eq:vacabjm} of the
algebraic R-matrix construction of invariants in ABJM theory is a
product of two-point invariants that might be the necessary starting
point to render the recursive symmetry in three dimensions exact, cf.\
the discussion in~\cite{Bargheer:2012cp}.
\medskip

As mentioned in \secref{sec:intro}, the study of deformed scattering
amplitudes in four dimensions was motivated by the map between the
one-loop dilatation operator and the four-point scattering amplitude
of $\superN=4$ SYM theory~\cite{Zwiebel:2011bx}. Construction of the
amplitude form of the associated R-matrix then led to the introduction
of a (spectral) deformation parameter~\cite{Ferro:2012xw,Ferro:2013dga}. In
this paper we have introduced
the deformation of scattering amplitudes in three dimensions. It would
be interesting to see how
the dilatation operator of ABJM theory can be constructed from
deformed amplitudes or on-shell diagrams.
\medskip

It would also be interesting to further explore the similarities with the
integrable structures discussed in the context of 4d $\superN=1$ quiver gauge
theories~\cite{Yamazaki:2012cp,Terashima:2012cx,Yamazaki:2013nra}.
Our discussion of ABJM scattering amplitudes suggests the existence of a new 3d
duality associated with the triangle move, cf.\ \eqref{eq:trianglemove}.
\medskip

Another notable question concerns the bonus symmetry of scattering
amplitudes in \mbox{$\superN=4$} SYM theory found in~\cite{Beisert:2011pn}.
Is this symmetry still preserved and what is its r\^ole for the
deformations of four-dimensional scattering amplitudes? Since the
generator of this level-one symmetry is bilocal in both the ordinary
and the dual conformal coordinates, and acts as a raising operator for
the Yangian levels, this might also clarify the relation between the
Yangian generators in twistor and momentum-twistor space discussed in
\secref{sec:defmomtwi}, which remains an interesting open problem. Finally,
studying these issues could shed light on the existence of a similar symmetry
in ABJM theory.
\medskip

Lastly, an important question is whether we can incorporate the above
deformations into other approaches like the amplituhedron
of~\cite{Arkani-Hamed:2013jha,Arkani-Hamed:2013kca}. Studying this question
would
be a good opportunity to elucidate the fate of Yangian symmetry in the
amplituhedron. Another recent development to use integrability for the
computation of scattering amplitudes is the non-perturbative flux-tube
formulation introduced in~\cite{Basso:2013vsa}. Since the approaches
of~\cite{Arkani-Hamed:2013jha,Arkani-Hamed:2013kca} and~\cite{Basso:2013vsa}
seem
to bring many advantages over the previous methods, combining them with the
deformation might be the most useful step in order to continue to investigate
the r\^ole of Yangian symmetry and the impact of integrability for amplitudes
in planar supersymmetric gauge theories.

\subsection*{Acknowledgments}

We are very grateful to Song He for numerous stimulating discussions
on the subject of this note and related topics, and for his initial collaboration.
We also thank Nima Arkani-Hamed, Sangmin Lee, and Carlo Meneghelli for interesting discussions.
The work of T.\,B.\ is supported by a Marie Curie International
Outgoing Fellowship within the 7$^\text{th}$ European Community Framework
Programme under Grant No.\ PIOF-GA-2011-299865.
The work of Y-t.\,H.\ is supported by the National Science Foundation
Grant No.\ PHY-1314311.
The work of F.\,L.\ was supported by a fellowship within the Postdoc-Program of
the German Academic Exchange Service (DAAD).
The work of M.\,Y.\ was supported in part by WPI program, MEXT, Japan.
He would also like to thank KITP/UCSB (``New Methods in Nonperturbative Quantum
Field Theory,'' NSF PHY11-25915) for hospitality where part of this work has
been performed.

\paragraph{Note added:} While this manuscript was in preparation, we found
out that the deformed Gra{\ss}mannian integrals for $\superN=4$ SYM theory
(both in momentum space and in momentum-twistor space) were
independently obtained by L.\,Ferro, T.\,{\L}ukowski and
M.\,Staudacher~\cite{Ferro:2014gca}.
We would like to thank them for correspondence, and for discussions during the
Strings 2014 conference.


\appendix
\addcontentsline{toc}{section}{Appendix}
\addtocontents{toc}{\protect\setcounter{tocdepth}{-1}}

\section{\texorpdfstring{Explicit $\alg{osp}(6|4)$}{osp(6|4)} Generators}
\label{sec:ABJMGen}

For reference,
here we list the level-zero generators of the $\alg{osp}(6|4)$ algebra in the
singleton representation as given in~\cite{Bargheer:2010hn}:
\begin{alignat*}{2}
\gen{L}^\alpha{}_{\beta}&=\lambda^\alpha\partial_\beta-\half\delta^\alpha_\beta\lambda^\gamma\partial_\gamma\,,&\qquad
\gen{P}^{\alpha\beta}&=\lambda^\alpha\lambda^\beta\,,\\
\gen{D}&=\half\lambda^\alpha\partial_\alpha+\half\,,&
\gen{K}_{\alpha\beta}&=\partial_\alpha\partial_\beta\,,
\end{alignat*}
\begin{equation*}
\gen{R}^{AB}=\eta^A\eta^B\,,\qquad
\gen{R}^A{}_B
=\eta^A\partial_B-\half\delta^A_B\,,\qquad
\gen{R}_{AB}=\partial_A\partial_B\,,
\end{equation*}
\begin{alignat}{2}
\gen{Q}^{\alpha A}&=\lambda^\alpha\eta^A\,,&\qquad
\gen{S}_\alpha^A&=\eta^A\partial_\alpha\,,\nn\\
\gen{Q}^\alpha_A&=\lambda^\alpha\partial_A\,,&
\gen{S}_{\alpha A}&=\partial_\alpha\partial_A\,.
\label{eq:genone}
\end{alignat}
See Appendices~F and~G of~\cite{Bargheer:2010hn} for the construction
of the level-one Yangian generators.

\section{Yangian Invariance of the 4d Deformed Gra{\ss}mannian Integral}
\label{app:4dgrassinv}

In this appendix, we check the Yangian invariance of
the deformed Gra{\ss}mannian formula \eqref{eq:topcell}.
The discussion is parallel to the case of the ABJM theory discussed in section \ref{sec:ABJM}.
We only need to prove the invariance under level-zero and level-one generators, since
all other generators can be obtained from the commutation relations.

The invariance under the level-zero is unaffected by the deformation:
It simply follows from the fact that the level-zero generators \eqref{eq:lev0} are
realized linearly in the twistor variables $\cZ_i$, and thus
annihilate the delta function $\delta^{4k|4k}(C\cdot\cZ)$
present in the Gra{\ss}mannian integral.

The invariance under the level-one generator will be verified
below following the methods of~\cite{Drummond:2010qh}.
Let us first rewrite $\gen{\widehat
J}^{\cA}{}_{\cB}$ as
\begin{align}
\gen{\widehat J}^{\cA}{}_{\cB}
=
\Bigbrk{2\sum_{i<j}-\sum_{i,j}+\sum_{i=j}}
\lrbrk{\cZ_i^\cA\frac{\partial}{\partial\cZ_j^\cB}\cZ_j^\cC
\frac{\partial}{\partial\cZ_i^\cC}-
\cZ_i^\cA\frac{\partial}{\partial\cZ_i^\cB}}
+\sum_i u_i\,\cZ_i^\cA\frac{\partial}{\partial\cZ_i^\cB}\,.
\end{align}
The sum $\sum_{i,j}$ gives a square of the level-zero generator and
acts trivially on the Gra{\ss}mannian formula. In the sum $\sum_{i=j}$
on the other hand, we find the central charge operator
$\gen{C}_i=-\cZ_i^\cC\,\partial/\partial\cZ_i^\cC$
\eqref{eq:centop}, which yields the following expression when acting on the
Gra{\ss}mannian integral
\begin{align}
2\sum_{i<j}\lrbrk{
\cZ_i^\cA\frac{\partial}{\partial
\cZ_j^\cB}\cZ_j^\cC\frac{\partial}{\partial\cZ_i^\cC}-
\cZ_i^\cA\frac{\partial}{\partial\cZ_i^\cB}
}
+\sum_i (u_i-c_i)\cZ_i^\cA\frac{\partial}{\partial\cZ_i^\cB}\,.
\end{align}
Now the crucial observation is that the operator
$\cZ_j^{\cC}\,\partial/\partial\cZ_i^{\cC}$, when acting on the delta functions,
can be replaced by a $\grp{GL}(k)$-rotation on the rows of the matrix~$C$~\cite{Drummond:2010qh}.

To do this properly, we need to fix the $\GL(k)$-gauge ambiguity as%
\footnote{For the deformed amplitude it is crucial to fix the $\GL(k)$-ambiguity to obtain
correct identification of deformation parameters. This contrasts with the case of the
undeformed case, where the formal analysis without fixing the
$\GL(k)$-ambiguity also
gives the Yangian invariance of the amplitude~\cite{Drummond:2010qh}.}
\begin{equation}
C=
\begin{pmatrix}
1 & \cdots & 0 & C_{1,k+1} & \cdots & C_{1,n} \\
0 & \cdots & 0 & \vdots & \vdots & \vdots \\
0 & \cdots & 1 & C_{k+1,n} & \cdots & C_{k,n}
\end{pmatrix}
\,.
\end{equation}
The operator $\cZ_j^{\cC}\,\partial/\partial\cZ_i^{\cC}$
then can be replaced by a $\grp{GL}(k)$-rotation on the row of the non-gauge-fixed part of the
matrix $C$.
It then follows that
\begin{align}
\gen{\widehat J}^{\cA}{}_\cB\grass_{k,n}
=
\sum_b\int\frac{\prod_{a=1}^k\prod_{m=k+1}^n dC_{am}}{M_1^{1+b_1}\dots M_n^{1+b_n}}
\lrsbrk{\cN_b^\cA-\cV_b^\cA+\cU{}_b^\cA}
(\partial_\cB\delta_b)\prod_{a\ne b}\delta_a\,,
\end{align}
where $\delta_a\equiv\delta^{4|4}\lrbrk{\cZ_a+\sum_{l=k+1}^n C_{al}\cZ_l}$,
\begin{align}
\begin{split}
\cN_b^\cA\equiv 2\sum_{i<j}\cN_{ij}\cZ_i^\cA C_{bj}\,,
\quad
\cV_b^\cA\equiv 2\sum_{i<j}\cZ_i^\cA C_{bi}\,,
\quad
\cU{}_b^\cA\equiv\sum_i u_i^{-}\cZ_i^\cA C_{bi}\,.
\end{split}
\end{align}
Here, $u_i^-=u_i-c_i$ as before,
and the operator $\cN_{ij}$ is a gauge-fixed version of the
operator $\sum_{a=1}^k C_{ai}\,\partial/\partial C_{aj}$~\cite{Drummond:2010qh}.

We can integrate by parts for the operator $\cN_b^A$.
The operator annihilates the measure, but
acts non-trivially on the minors $M_i(C)$.
Generalizing the commutation relations of~\cite{Drummond:2010qh},
we find
\begin{align}
\lrcomm{\frac{1}{M_1^{1+b_1}\dots M_n^{1+b_n}}}{\cN_b^\cA}
=
\frac{1}{M_1^{1+b_1}\dots M_n^{1+b_n}}
\sum_{i<j} (1+b_j) Z_i^\cA C_{bi}\,,
\end{align}
and therefore
\begin{align}
\gen{\widehat J}^{\cA}{}_\cB\cA_{k,n}
=\int\frac{\prod_{a=1}^k\prod_{m=k+1}^n dt_{am}}{M_1^{1+b_1}\dots M_n^{1+b_n}}
\sum_b\Bigsbrk{2\sum_{i<j}b_j Z_i^\cA C_{bi}+\sum_i u_i^{-}Z_i^\cA C_{bi}}
(\partial_B\delta_b)\prod_{a\ne b}\delta_a\,.
\end{align}
Requiring that the coefficient of $Z_i^A C_{bi}$ is a constant, we thus find
the invariance constraints
\begin{align}
\sum_{j=i+1}^n 2b_j+u_i^{-}=\text{const}
\label{eq:cons4d}
\end{align}
for all $i=1,\dots,n$, and where the constant is independent of $i$.
In other words we have
\begin{align}
\label{eq:uc}
b_i=\half(u_i^{-}-u_{i-1}^{-})\,,
\end{align}
which agrees with the constraints in \eqref{eq:topcellexp}.

\section{Yangian Invariance of the 3d Gra{\ss}mannian Integral}
\label{sec:Grass}

In this appendix, we prove the Yangian invariance of the (undeformed)
orthogonal Gra{\ss}mannian integral:
\begin{equation}
\grass_{2k}=\int\frac{d^{k\times 2k} C}{|\GL(k)|}
\frac{\delta^{k(k+1)/2}(C\cdot C\transpose)\,\delta^{2k|3k}(C\cdot\Lambda)}
{\prod_{i=1}^{k} M_i(C)}\,.
\end{equation}
To show that the above integral is invariant under the level-one
generator in \eqref{J1Def1}, we begin by rewriting again
\begin{align}
(-)^{|\cC|}
\Lambda^{(\cA}_l
\frac{\partial}{\partial\Lambda_l^{\cC}}
\Lambda_i^{\cC}\Lambda^{\cB]}_i
=
\Lambda^{(\cA}_l
\Lambda^{\cB]}_i
\Lambda_i^{\cC}
\frac{\partial}{\partial\Lambda_l^{\cC}}
\equiv
\Lambda^{(\cA}_l\Lambda^{\cB]}_i\cO_i\,^l\,,
\end{align}
where $\cO_i\,^l$ is simply a $\GL(2k)$-rotation on the
external data $\Lambda_i$, and we again can conveniently rewrite the
action of the first term in \eqref{J1Def1} as
\begin{equation}
\label{ManipulationO1}
\sum_{l<i}
\lrbrk{
\Lambda^{(\cA}_l
\Lambda^{\cB]}_i
\cO_i\,^l
-
\Lambda^{(\cA}_i
\Lambda^{\cB]}_l
\cO_l\,^i
}
\delta^{2|3}(C\cdot\Lambda)
=
\sum_{l<i}
\Lambda^{(\cA}_l\Lambda^{\cB]}_i
\bigbrk{O_l\,^i-O_i\,^l}\,
\delta^{2|3}(C\cdot\Lambda)\,,
\end{equation}
where $O_i\,^l $ is defined in \eqref{ODef}, and in obtaining the last
line we have used the fact that the indices of the level-one
generators under consideration are (anti-)symmetrized. Since the
operator in the square bracket is in fact an
$\mathrm{O}(2k)$-rotation, after integration by parts it vanishes when
acting on the
$\mathrm{O}(2k)$ invariant constraint $\delta(C\cdot C\transpose)$.
Thus the only contribution we receive after integration by parts is
when the linear operator acts on the minors:
\begin{align}
\sum_{l<i}
O_i\,^l M_{p}
=
\sum_{l=p}^{p+k-1}\sum_{i=p+k}^{2k}
M_{p}^{l\rightarrow i}\,,
\quad
\sum_{l<i}
O_l\,^i M_{p}
=
\sum_{l=1}^{p-1}\sum_{i=p}^{p+k-1}
M_{p}^{i\rightarrow l}\,.
\end{align}
Finally, since on the support of $\delta(C\cdot C\transpose)$, the matrix $C$
is a collection of null $k$-planes in a $2k$-dimensional space, one can define
a set of dual $k$-planes to construct $\hat{C}$ such that~\cite{Lee:2010du}
\begin{align}
\label{CC}
\hat{C}\cdot\hat{C}\transpose=0\,,
\quad
C\cdot\hat{C}\transpose=\hat{C}\cdot C\transpose=I_{k\times k}\,.
\end{align}
Note that due to \eqref{CC}, one can immediately deduce
\begin{align}
\label{Ikk}
C\transpose\cdot\hat{C}+\hat{C}\transpose\cdot C=I_{2k\times 2k}\,.
\end{align}
This is a useful identity, since we can now rewrite
\begin{align}
\Lambda^{\cA}_i
=
\sum_{j=1}^{2k}
\Lambda^{\cA}_j
\sum_{a}(C_{ja}\hat{C}_{ia}+\hat{C}_{ja}C_{ia})\,.
\end{align}
On the support of $\delta^{2|3}(C\cdot\Lambda)$, the first term vanishes. Using
this result, with $p\leq k$, we find that
\begin{multline}
\sum_{l<i}
\Lambda^{(\cA}_i
\Lambda^{\cB]}_l
O_i\,^l M_{p}
=
\sum_{l<i}\sum_{j=1}^{2k}
\Lambda^{(\cA}_i
\Lambda^{\cB]}_j
\sum_{a}(\hat{C}_{ja}C_{la})O_i\,^l M_{p}
\\
=
\sum_{j=1}^{2k}\sum_{i=p+k}^{2k}
\Lambda^{(\cA}_i
\Lambda^{\cB]}_j
\sum_{a}\hat{C}_{ja}\sum_{l=p}^{p+k-1}
M_{p}^{l\rightarrow i}\,
C_{la}
=
\sum_{i=p+k}^{2k}
\Lambda^{(\cA}_i
\Lambda^{\cB]}_i
M_{p}\,,
\end{multline}
where a $k$-term Schouten identity was used in the last line, as well
as the completeness relation in \eqref{Ikk}. This leads to the
following rewriting of the first term in \eqref{ManipulationO1}:
\begin{equation}
\label{il}
\sum_{l<i}\Lambda^{(\cA}_i\Lambda^{\cB]}_l O_i\,^l\frac{1}{\prod_{j=1}^k M_j}
=
-\frac{\lrbrk{\sum_{l=1}^{k}\sum_{i=l+k}^{2k}\Lambda^{(\cA}_i\Lambda^{\cB]}_i}}{\prod_{j=1}^k M_j}
=
\frac{1}{\prod_{j=1}^k M_j}
\sum_{k\leq l<i}\Lambda^{(\cA}_i\Lambda^{\cB]}_i
\,.
\end{equation}
Similarly, we find:
\begin{multline}
\sum_{l<i}\Lambda^{(\cA}_i\Lambda^{\cB]}_l\,O_l\,^i
M_{p}
=
\sum_{l<i}\sum_{j=1}^{2k}
\Lambda^{(\cA}_j\Lambda^{\cB]}_l\sum_{a}\hat{C}_{aj}C_{ai}\,O_l\,^i M_{p}
\\
=
\sum_{l=1}^{p-1}\sum_{j=1}^{2k}\Lambda^{(\cA}_j\Lambda^{\cB]}_l
\sum_{a}\hat{C}_{aj}\sum_{i=p}^{p+k-1}C_{ai}\,M_{p}^{i\rightarrow l}
=
\sum_{l=1}^{p-1}\Lambda^{(\cA}_l\Lambda^{\cB]}_lM_{p}\,.
\end{multline}
Hence, for the second term in \eqref{ManipulationO1} we now have
\begin{equation}
\label{li}
-\sum_{l<i}
\Lambda^{(\cA}_i
\Lambda^{\cB]}_l
O_l\,^i
\frac{1}{\prod_{j=1}^k M_j}
=
\frac{1}{\prod_{j=1}^k M_j}
\sum_{l<i\leq k}
\Lambda^{(\cA}_l
\Lambda^{\cB]}_l\,.
\end{equation}
Collecting these results, and using the level-zero
constraint $\sum_i\Lambda_i^{\cA}\Lambda_i^{\cB}=0$,
we find that \eqref{il} and \eqref{li} is
exactly what is needed to cancel against the terms of the form
$\Lambda_i\Lambda_i$ in \eqref{J1Def1}. For example, for $n=2k=4$,
the $\Lambda_i\Lambda_i$ terms in \eqref{J1Def1} are given by
\begin{align}
\label{Anomaly}
\biggbrk{\sum_{l<i}-\sum_{i<l}}
\frac{\Lambda^{(\cA}_i\Lambda^{\cB]}_i}{2}
&=
\frac{1}{2}
\lrbrk{
-3\Lambda^{(\cA}_1\Lambda^{\cB]}_1
-\Lambda^{(\cA}_2\Lambda^{\cB]}_2
+\Lambda^{(\cA}_3\Lambda^{\cB]}_3
+3\Lambda^{(\cA}_4\Lambda^{\cB]}_4}\nn\\
&=
-3\Lambda^{(\cA}_1\Lambda^{\cB]}_1
-2\Lambda^{(\cA}_2\Lambda^{\cB]}_2
-\Lambda^{(\cA}_3\Lambda^{\cB]}_3\,.
\end{align}
On the other hand \eqref{il} and \eqref{li} yield
\begin{equation}
\lrbrk{
-\Lambda^{(\cA}_3\Lambda^{\cB]}_3
-2\Lambda^{(\cA}_4\Lambda^{\cB]}_4
}
+\brk{\Lambda^{(\cA}_1\Lambda^{\cB]}_1}
=
3\Lambda^{(\cA}_1\Lambda^{\cB]}_1
+2\Lambda^{(\cA}_2\Lambda^{\cB]}_2
+\Lambda^{(\cA}_3\Lambda^{\cB]}_3\,.
\end{equation}
Indeed the above is what is necessary to cancel \eqref{Anomaly}. This completes
the proof of the invariance of the orthogonal Gra{\ss}mannian integral under
the level-one generator in \eqref{J1Def1}.

\pdfbookmark[1]{\refname}{references}
\bibliographystyle{nb}
\bibliography{IntAmpDef}

\begin{thebibliography}{10}
\providecommand{\href}[2]{#2}
\providecommand{\arxivref}[2]{\href{http://arxiv.org/abs/#1}{#2}}
\providecommand{\doiref}[2]{\href{http://dx.doi.org/#1}{#2}}
\providecommand{\nbbstauthor}[1]{#1}
\providecommand{\nbbstjournal}[1]{\textsf{#1}}
\providecommand{\nbbsttitle}[1]{\textit{#1}}
\providecommand{\nbbsturl}[1]{\texttt{#1}}
\providecommand{\nbbsteprint}[1]{\texttt{#1}}
\providecommand{\nbbststyle}{\raggedright\small\parskip0pt}
\nbbststyle

\bibitem{Elvang:2013cua}
\nbbstauthor{H.~Elvang and Y.-t.~Huang},
\nbbsttitle{``Scattering Amplitudes''},
\nbbsteprint{\arxivref{1308.1697}{arxiv:1308.1697}}.

\bibitem{Aharony:2008ug}
\nbbstauthor{O.~Aharony, O.~Bergman, D.~L.~Jafferis and J.~Maldacena},
\nbbsttitle{``{$\mathcal{N}=\mathord{}$6} superconformal Chern--Simons-matter
  theories, M2-branes and their gravity duals''},
\nbbstjournal{\doiref{10.1088/1126-6708/2008/10/091}{JHEP~0810,~091~(2008)}},
\nbbsteprint{\arxivref{0806.1218}{arxiv:0806.1218}}.

\bibitem{Hosomichi:2008jb}
\nbbstauthor{K.~Hosomichi, K.-M.~Lee, S.~Lee, S.~Lee and J.~Park},
\nbbsttitle{``{$\mathcal{N}=\mathord{}$5,6} Superconformal Chern--Simons
  Theories and M2-branes on Orbifolds''},
\nbbstjournal{\doiref{10.1088/1126-6708/2008/09/002}{JHEP~0809,~002~(2008)}},
\nbbsteprint{\arxivref{0806.4977}{arxiv:0806.4977}}.

\bibitem{Drummond:2009fd}
\nbbstauthor{J.~M.~Drummond, J.~M.~Henn and J.~Plefka},
\nbbsttitle{``Yangian symmetry of scattering amplitudes in
  {$\mathcal{N}=\mathord{}$4} super Yang--Mills theory''},
\nbbstjournal{\doiref{10.1088/1126-6708/2009/05/046}{JHEP~0905,~046~(2009)}},
\nbbsteprint{\arxivref{0902.2987}{arxiv:0902.2987}}.

\bibitem{Bargheer:2009qu}
\nbbstauthor{T.~Bargheer, N.~Beisert, W.~Galleas, F.~Loebbert and
  T.~McLoughlin},
\nbbsttitle{``Exacting {$\mathcal{N}=\mathord{}$4} Superconformal Symmetry''},
\nbbstjournal{\doiref{10.1088/1126-6708/2009/11/056}{JHEP~0911,~056~(2009)}},
\nbbsteprint{\arxivref{0905.3738}{arxiv:0905.3738}}.

\bibitem{Korchemsky:2009hm}
\nbbstauthor{G.~P.~Korchemsky and E.~Sokatchev},
\nbbsttitle{``Symmetries and analytic properties of scattering amplitudes in
  {$\mathcal{N}=\mathord{}$4} SYM theory''},
\nbbstjournal{\doiref{10.1016/j.nuclphysb.2010.01.022}{Nucl.~Phys.~B832,~1~(2010)}},
\nbbsteprint{\arxivref{0906.1737}{arxiv:0906.1737}}.

\bibitem{Zwiebel:2011bx}
\nbbstauthor{B.~I.~Zwiebel},
\nbbsttitle{``From Scattering Amplitudes to the Dilatation Generator in
  {$\mathcal{N}=\mathord{}$4} SYM''},
\nbbstjournal{\doiref{10.1088/1751-8113/45/11/115401}{J.~Phys~A45,~115401~(2012)}},
\nbbsteprint{\arxivref{1111.0083}{arxiv:1111.0083}}.

\bibitem{Ferro:2012xw}
\nbbstauthor{L.~Ferro, T.~{\L}ukowski, C.~Meneghelli, J.~Plefka and
  M.~Staudacher},
\nbbsttitle{``Harmonic R-matrices for Scattering Amplitudes and Spectral
  Regularization''},
\nbbstjournal{\doiref{10.1103/PhysRevLett.110.121602}{Phys.~Rev.~Lett.~110,~121602~(2013)}},
\nbbsteprint{\arxivref{1212.0850}{arxiv:1212.0850}}.

\bibitem{Ferro:2013dga}
\nbbstauthor{L.~Ferro, T.~{\L}ukowski, C.~Meneghelli, J.~Plefka and
  M.~Staudacher},
\nbbsttitle{``Spectral Parameters for Scattering Amplitudes in
  {$\mathcal{N}=\mathord{}$4} Super Yang--Mills Theory''},
\nbbstjournal{\doiref{10.1007/JHEP01(2014)094}{JHEP~1401,~094~(2014)}},
\nbbsteprint{\arxivref{1308.3494}{arxiv:1308.3494}}.

\bibitem{ArkaniHamed:2012nw}
\nbbstauthor{N.~Arkani-Hamed, J.~L.~Bourjaily, F.~Cachazo, A.~B.~Goncharov,
  A.~Postnikov and J.~Trnka},
\nbbsttitle{``Scattering Amplitudes and the Positive Grassmannian''},
\nbbsteprint{\arxivref{1212.5605}{arxiv:1212.5605}}.

\bibitem{Beisert:2014qba}
\nbbstauthor{N.~Beisert, J.~Broedel and M.~Rosso},
\nbbsttitle{``On Yangian-invariant regularization of deformed on-shell diagrams
  in {$\mathcal{N}=\mathord{}$4} super-Yang--Mills theory''},
\nbbstjournal{\doiref{10.1088/1751-8113/47/36/365402}{J.~Phys.~A47,~365402~(2014)}},
\nbbsteprint{\arxivref{1401.7274}{arxiv:1401.7274}}.

\bibitem{Chicherin:2013ora}
\nbbstauthor{D.~Chicherin, S.~Derkachov and R.~Kirschner},
\nbbsttitle{``Yang--Baxter operators and scattering amplitudes in
  {$\mathcal{N}=\mathord{}$4} super-Yang--Mills theory''},
\nbbstjournal{\doiref{10.1016/j.nuclphysb.2014.02.016}{Nucl.~Phys.~B881,~467~(2014)}},
\nbbsteprint{\arxivref{1309.5748}{arxiv:1309.5748}}.

\bibitem{Frassek:2013xza}
\nbbstauthor{R.~Frassek, N.~Kanning, Y.~Ko and M.~Staudacher},
\nbbsttitle{``Bethe Ansatz for Yangian Invariants: Towards Super Yang--Mills
  Scattering Amplitudes''},
\nbbstjournal{\doiref{10.1016/j.nuclphysb.2014.03.015}{Nucl.~Phys.~B883,~373~(2014)}},
\nbbsteprint{\arxivref{1312.1693}{arxiv:1312.1693}}.

\bibitem{Kanning:2014maa}
\nbbstauthor{N.~Kanning, T.~{\L}ukowski and M.~Staudacher},
\nbbsttitle{``A Shortcut to General Tree-level Scattering Amplitudes in
  {$\mathcal{N}=\mathord{}$4} SYM via Integrability''},
\nbbstjournal{\doiref{10.1002/prop.201400017}{Fortsch.~Phys.~62,~556~(2014)}},
\nbbsteprint{\arxivref{1403.3382}{arxiv:1403.3382}}.

\bibitem{Broedel:2014pia}
\nbbstauthor{J.~Broedel, M.~de~Leeuw and M.~Rosso},
\nbbsttitle{``A dictionary between $R$-operators, on-shell graphs and Yangian
  algebras''},
\nbbstjournal{\doiref{10.1007/JHEP06(2014)170}{JHEP~1406,~170~(2014)}},
\nbbsteprint{\arxivref{1403.3670}{arxiv:1403.3670}}.

\bibitem{Broedel:2014hca}
\nbbstauthor{J.~Broedel, M.~de~Leeuw and M.~Rosso},
\nbbsttitle{``Deformed one-loop amplitudes in {$\mathcal{N}=\mathord{}$4}
  super-Yang--Mills theory''},
\nbbstjournal{\doiref{10.1007/JHEP11(2014)091}{JHEP~1411,~091~(2014)}},
\nbbsteprint{\arxivref{1406.4024}{arxiv:1406.4024}}.

\bibitem{Britto:2004ap}
\nbbstauthor{R.~Britto, F.~Cachazo and B.~Feng},
\nbbsttitle{``New recursion relations for tree amplitudes of gluons''},
\nbbstjournal{\doiref{10.1016/j.nuclphysb.2005.02.030}{Nucl.~Phys.~B715,~499~(2005)}},
\nbbsteprint{\arxivref{hep-th/0412308}{hep-th/0412308}}.

\bibitem{Britto:2005fq}
\nbbstauthor{R.~Britto, F.~Cachazo, B.~Feng and E.~Witten},
\nbbsttitle{``Direct proof of tree-level recursion relation in Yang--Mills
  theory''},
\nbbstjournal{\doiref{10.1103/PhysRevLett.94.181602}{Phys.~Rev.~Lett.~94,~181602~(2005)}},
\nbbsteprint{\arxivref{hep-th/0501052}{hep-th/0501052}}.

\bibitem{ArkaniHamed:2009dn}
\nbbstauthor{N.~Arkani-Hamed, F.~Cachazo, C.~Cheung and J.~Kaplan},
\nbbsttitle{``A Duality For The S~Matrix''},
\nbbstjournal{\doiref{10.1007/JHEP03(2010)020}{JHEP~1003,~020~(2010)}},
\nbbsteprint{\arxivref{0907.5418}{arxiv:0907.5418}}.

\bibitem{Huang:2013owa}
\nbbstauthor{Y.-T.~Huang and C.~Wen},
\nbbsttitle{``ABJM amplitudes and the positive orthogonal grassmannian''},
\nbbstjournal{\doiref{10.1007/JHEP02(2014)104}{JHEP~1402,~104~(2014)}},
\nbbsteprint{\arxivref{1309.3252}{arxiv:1309.3252}}.

\bibitem{Lee:2010du}
\nbbstauthor{S.~Lee},
\nbbsttitle{``Yangian Invariant Scattering Amplitudes in Supersymmetric
  Chern--Simons Theory''},
\nbbstjournal{\doiref{10.1103/PhysRevLett.105.151603}{Phys.~Rev.~Lett.~105,~151603~(2010)}},
\nbbsteprint{\arxivref{1007.4772}{arxiv:1007.4772}}.

\bibitem{Bargheer:2010hn}
\nbbstauthor{T.~Bargheer, F.~Loebbert and C.~Meneghelli},
\nbbsttitle{``Symmetries of Tree-level Scattering Amplitudes in
  {$\mathcal{N}=\mathord{}$6} Superconformal Chern--Simons Theory''},
\nbbstjournal{\doiref{10.1103/PhysRevD.82.045016}{Phys.~Rev.~D82,~045016~(2010)}},
\nbbsteprint{\arxivref{1003.6120}{arxiv:1003.6120}}.

\bibitem{Beisert:2010jq}
\nbbstauthor{N.~Beisert},
\nbbsttitle{``On Yangian Symmetry in Planar {$\mathcal{N}=\mathord{}$4} SYM''},
\nbbsteprint{\arxivref{1004.5423}{arxiv:1004.5423}},
in: \nbbsttitle{``Quantum Chromodynamics and Beyond: Gribov-80 Memorial
  Volume''},
ed.: Y.~L.~Dokshitzer, P.~L\'evai and J.~Ny\'iri,
World Scientific (2011),
Singapore.

\bibitem{Bernard:1993ya}
\nbbstauthor{D.~Bernard},
\nbbsttitle{``An Introduction to Yangian Symmetries''},
\nbbstjournal{\doiref{10.1142/S0217979293003371}{Int.~J.~Mod.~Phys.~B07,~3517~(1993)}},
\nbbsteprint{\arxivref{hep-th/9211133}{hep-th/9211133}}.

\bibitem{MacKay:2004tc}
\nbbstauthor{N.~J.~MacKay},
\nbbsttitle{``Introduction to Yangian symmetry in integrable field theory''},
\nbbstjournal{\doiref{10.1142/S0217751X05022317}{Int.~J.~Mod.~Phys.~A20,~7189~(2005)}},
\nbbsteprint{\arxivref{hep-th/0409183}{hep-th/0409183}}.

\bibitem{Witten:2003nn}
\nbbstauthor{E.~Witten},
\nbbsttitle{``Perturbative gauge theory as a string theory in twistor space''},
\nbbstjournal{\doiref{10.1007/s00220-004-1187-3}{Commun.~Math.~Phys.~252,~189~(2004)}},
\nbbsteprint{\arxivref{hep-th/0312171}{hep-th/0312171}}.

\bibitem{Nair:1988bq}
\nbbstauthor{V.~P.~Nair},
\nbbsttitle{``A current algebra for some gauge theory amplitudes''},
\nbbstjournal{\doiref{10.1016/0370-2693(88)91471-2}{Phys.~Lett.~B214,~215~(1988)}}.

\bibitem{Ferro:2014gca}
\nbbstauthor{L.~Ferro, T.~{\L}ukowski and M.~Staudacher},
\nbbsttitle{``{$\mathcal{N}=\mathord{}$4} scattering amplitudes and the
  deformed Gra{\ss}mannian''},
\nbbstjournal{\doiref{10.1016/j.nuclphysb.2014.10.012}{Nucl.~Phys.~B889,~192~(2014)}},
\nbbsteprint{\arxivref{1407.6736}{arxiv:1407.6736}}.

\bibitem{Drummond:2010qh}
\nbbstauthor{J.~M.~Drummond and L.~Ferro},
\nbbsttitle{``Yangians, Grassmannians and T-duality''},
\nbbstjournal{\doiref{10.1007/JHEP07(2010)027}{JHEP~1007,~027~(2010)}},
\nbbsteprint{\arxivref{1001.3348}{arxiv:1001.3348}}.

\bibitem{Mason:2009qx}
\nbbstauthor{L.~Mason and D.~Skinner},
\nbbsttitle{``Dual Superconformal Invariance, Momentum Twistors and
  Grassmannians''},
\nbbstjournal{\doiref{10.1088/1126-6708/2009/11/045}{JHEP~0911,~045~(2009)}},
\nbbsteprint{\arxivref{0909.0250}{arxiv:0909.0250}}.

\bibitem{Bourjaily:2012gy}
\nbbstauthor{J.~L.~Bourjaily},
\nbbsttitle{``Positroids, Plabic Graphs, and Scattering Amplitudes in
  Mathematica''},
\nbbsteprint{\arxivref{1212.6974}{arxiv:1212.6974}}.

\bibitem{Huang:2014xza}
\nbbstauthor{Y.-t.~Huang, C.~Wen and D.~Xie},
\nbbsttitle{``The Positive orthogonal Grassmannian and loop amplitudes of
  ABJM''},
\nbbstjournal{\doiref{10.1088/1751-8113/47/47/474008}{J.~Phys.~A47,~474008~(2014)}},
\nbbsteprint{\arxivref{1402.1479}{arxiv:1402.1479}}.

\bibitem{Drummond:2010uq}
\nbbstauthor{J.~M.~Drummond and L.~Ferro},
\nbbsttitle{``The Yangian origin of the Grassmannian integral''},
\nbbstjournal{\doiref{10.1007/JHEP12(2010)010}{JHEP~1012,~010~(2010)}},
\nbbsteprint{\arxivref{1002.4622}{arxiv:1002.4622}}.

\bibitem{Bargheer:2012cp}
\nbbstauthor{T.~Bargheer, N.~Beisert, F.~Loebbert and T.~McLoughlin},
\nbbsttitle{``Conformal Anomaly for Amplitudes in {$\mathcal{N}=\mathord{}$6}
  Superconformal Chern--Simons Theory''},
\nbbstjournal{\doiref{10.1088/1751-8113/45/47/475402}{J.~Phys.~A45,~475402~(2012)}},
\nbbsteprint{\arxivref{1204.4406}{arxiv:1204.4406}}.

\bibitem{Gang:2010gy}
\nbbstauthor{D.~Gang, Y.-t.~Huang, E.~Koh, S.~Lee and A.~E.~Lipstein},
\nbbsttitle{``Tree-level Recursion Relation and Dual Superconformal Symmetry of
  the ABJM Theory''},
\nbbstjournal{\doiref{10.1007/JHEP03(2011)116}{JHEP~1103,~116~(2011)}},
\nbbsteprint{\arxivref{1012.5032}{arxiv:1012.5032}}.

\bibitem{Dolan:2004ys}
\nbbstauthor{L.~Dolan and C.~R.~Nappi},
\nbbsttitle{``Spin models and superconformal Yang--Mills theory''},
\nbbstjournal{\doiref{10.1016/j.nuclphysb.2005.04.006}{Nucl.~Phys.~B717,~361~(2005)}},
\nbbsteprint{\arxivref{hep-th/0411020}{hep-th/0411020}}.

\bibitem{Dolan:2004ps}
\nbbstauthor{L.~Dolan, C.~R.~Nappi and E.~Witten},
\nbbsttitle{``Yangian symmetry in $D=$4 superconformal Yang--Mills theory''},
\nbbsteprint{\arxivref{hep-th/0401243}{hep-th/0401243}},
in: \nbbsttitle{``Quantum Theory and Symmetries''},
ed.: P.~C.~Argyres et~al.,
World Scientific (2004),
Singapore.

\bibitem{Takhtajan:1979iv}
\nbbstauthor{L.~A.~Takhtajan and L.~D.~Faddeev},
\nbbsttitle{``The Quantum method of the inverse problem and the Heisenberg XYZ
  model''},
\nbbstjournal{Russ.~Math.~Surveys~34,~11~(1979)}.

\bibitem{Kulish:1981gi}
\nbbstauthor{P.~P.~Kulish, N.~{\relax Yu}.~Reshetikhin and E.~K.~Sklyanin},
\nbbsttitle{``Yang--Baxter equation and representation theory: I''},
\nbbstjournal{Lett.~Math.~Phys.~5,~393~(1981)}.

\bibitem{Kulish:1981bi}
\nbbstauthor{P.~P.~Kulish and E.~K.~Sklyanin},
\nbbsttitle{``Quantum Spectral Transform Method. Recent Developments''},
\nbbstjournal{Lect.~Notes~Phys.~151,~61~(1982)}.

\bibitem{Faddeev:1996iy}
\nbbstauthor{L.~D.~Faddeev},
\nbbsttitle{``How Algebraic Bethe Ansatz works for integrable model''},
\nbbsteprint{\arxivref{hep-th/9605187}{hep-th/9605187}},
in: \nbbsttitle{``Relativistic gravitation and gravitational radiation''},
ed.: J.-A.~Marck and J.-P.~Lasota,
Cambridge University Press (1997),
Cambridge.

\bibitem{Drummond:2008vq}
\nbbstauthor{J.~M.~Drummond, J.~Henn, G.~P.~Korchemsky and E.~Sokatchev},
\nbbsttitle{``Dual superconformal symmetry of scattering amplitudes in
  {$\mathcal{N}=\mathord{}$4} super-Yang--Mills theory''},
\nbbstjournal{\doiref{10.1016/j.nuclphysb.2009.11.022}{Nucl.~Phys.~B828,~317~(2010)}},
\nbbsteprint{\arxivref{0807.1095}{arxiv:0807.1095}}.

\bibitem{Schuster:2013pxj}
\nbbstauthor{P.~Schuster and N.~Toro},
\nbbsttitle{``On the Theory of Continuous-Spin Particles: Wavefunctions and
  Soft-Factor Scattering Amplitudes''},
\nbbstjournal{\doiref{10.1007/JHEP09(2013)104}{JHEP~1309,~104~(2013)}},
\nbbsteprint{\arxivref{1302.1198}{arxiv:1302.1198}}.

\bibitem{Schuster:2014xja}
\nbbstauthor{P.~Schuster and N.~Toro},
\nbbsttitle{``A New Class of Particle in 2+1 Dimensions''},
\nbbsteprint{\arxivref{1404.1076}{arxiv:1404.1076}}.

\bibitem{Beisert:2010gn}
\nbbstauthor{N.~Beisert, J.~Henn, T.~McLoughlin and J.~Plefka},
\nbbsttitle{``One-Loop Superconformal and Yangian Symmetries of Scattering
  Amplitudes in {$\mathcal{N}=\mathord{}$4} Super Yang--Mills''},
\nbbstjournal{\doiref{10.1007/JHEP04(2010)085}{JHEP~1004,~085~(2010)}},
\nbbsteprint{\arxivref{1002.1733}{arxiv:1002.1733}}.

\bibitem{Bargheer:2011mm}
\nbbstauthor{T.~Bargheer, N.~Beisert and F.~Loebbert},
\nbbsttitle{``Exact Superconformal and Yangian Symmetry of Scattering
  Amplitudes''},
\nbbstjournal{\doiref{10.1088/1751-8113/44/45/454012}{J.~Phys.~A44,~454012~(2011)}},
\nbbsteprint{\arxivref{1104.0700}{arxiv:1104.0700}}.

\bibitem{Bianchi:2012cq}
\nbbstauthor{M.~S.~Bianchi, M.~Leoni, A.~Mauri, S.~Penati and A.~Santambrogio},
\nbbsttitle{``One Loop Amplitudes In ABJM''},
\nbbstjournal{\doiref{10.1007/JHEP07(2012)029}{JHEP~1207,~029~(2012)}},
\nbbsteprint{\arxivref{1204.4407}{arxiv:1204.4407}}.

\bibitem{Yamazaki:2012cp}
\nbbstauthor{M.~Yamazaki},
\nbbsttitle{``Quivers, YBE and 3-manifolds''},
\nbbstjournal{\doiref{10.1007/JHEP05(2012)147}{JHEP~1205,~147~(2012)}},
\nbbsteprint{\arxivref{1203.5784}{arxiv:1203.5784}}.

\bibitem{Terashima:2012cx}
\nbbstauthor{Y.~Terashima and M.~Yamazaki},
\nbbsttitle{``Emergent 3-manifolds from 4d Superconformal Indices''},
\nbbstjournal{\doiref{10.1103/PhysRevLett.109.091602}{Phys.~Rev.~Lett.~109,~091602~(2012)}},
\nbbsteprint{\arxivref{1203.5792}{arxiv:1203.5792}}.

\bibitem{Yamazaki:2013nra}
\nbbstauthor{M.~Yamazaki},
\nbbsttitle{``New Integrable Models from the Gauge/YBE Correspondence''},
\nbbstjournal{\doiref{10.1007/s10955-013-0884-8}{J.~Statist.~Phys.~154,~895~(2014)}},
\nbbsteprint{\arxivref{1307.1128}{arxiv:1307.1128}}.

\bibitem{Beisert:2011pn}
\nbbstauthor{N.~Beisert and B.~U.~W.~Schwab},
\nbbsttitle{``Bonus Yangian Symmetry for the Planar S-Matrix of
  {$\mathcal{N}=\mathord{}$4} Super Yang--Mills''},
\nbbstjournal{Phys.~Rev.~Lett.~106,~231602~(2011)},
\nbbsteprint{\arxivref{1103.0646}{arxiv:1103.0646}}.

\bibitem{Arkani-Hamed:2013jha}
\nbbstauthor{N.~Arkani-Hamed and J.~Trnka},
\nbbsttitle{``The Amplituhedron''},
\nbbstjournal{\doiref{10.1007/JHEP10(2014)030}{JHEP~1410,~30~(2014)}},
\nbbsteprint{\arxivref{1312.2007}{arxiv:1312.2007}}.

\bibitem{Arkani-Hamed:2013kca}
\nbbstauthor{N.~Arkani-Hamed and J.~Trnka},
\nbbsttitle{``Into the Amplituhedron''},
\nbbsteprint{\arxivref{1312.7878}{arxiv:1312.7878}}.

\bibitem{Basso:2013vsa}
\nbbstauthor{B.~Basso, A.~Sever and P.~Vieira},
\nbbsttitle{``Spacetime and Flux Tube S-Matrices at Finite Coupling for
  {$\mathcal{N}=\mathord{}$4} Supersymmetric Yang--Mills Theory''},
\nbbstjournal{\doiref{10.1103/PhysRevLett.111.091602}{Phys.~Rev.~Lett.~111,~091602~(2013)}},
\nbbsteprint{\arxivref{1303.1396}{arxiv:1303.1396}}.

\end{thebibliography}

\end{document}